\documentclass[11pt,letterpaper,preprint]{article}
\usepackage{adjustbox}
\usepackage{booktabs}
\usepackage[utf8]{inputenc}
\usepackage{amsmath}
\usepackage{amssymb}
\usepackage[english]{babel}
\usepackage{graphics}
\usepackage{graphicx}
\usepackage{epsfig}
\usepackage{verbatim}
\usepackage{latexsym}
\usepackage{color}
\usepackage{accents}
\usepackage{float}
\usepackage{caption}
\usepackage{subcaption}
\usepackage{rotating}
\usepackage[T1]{fontenc}
\usepackage{ulem}
\usepackage[margin=1in]{geometry}
\usepackage{authblk}

\begin{document}
\title{A cluster driven log-volatility factor model: a deepening on the source of the volatility clustering}

\author[1]{Anshul Verma\thanks{Corresponding author, anshul.verma@kcl.ac.uk, +447740779724}}
\author[1]{R. J. Buonocore\thanks{riccardo\_junior.buonocore@kcl.ac.uk,+447549919717}}
\author[1,2,3]{T. Di Matteo\thanks{tiziana.di\_matteo@kcl.ac.uk,+4402078482223}}

\affil[1]{Department of Mathematics, King's College London, The Strand, London, WC2R 2LS, UK}
\affil[2]{Department of Computer Science, University College London, Gower Street, London, WC1E 6BT, UK}
\affil[3]{Complexity Science Hub Vienna, Josefstaedter Strasse 39, A 1080 Vienna}
\maketitle
\begin{abstract}
We introduce a new factor model for log volatilities that performs dimensionality reduction and considers contributions globally through the market, and locally through cluster structure and their interactions. We do not assume a-priori the number of clusters in the data, instead using the Directed Bubble Hierarchical Tree (DBHT) algorithm to fix the number of factors. We use the factor model and a new integrated non parametric proxy to study how volatilities contribute to volatility clustering. Globally, only the market contributes to the volatility clustering. Locally for some clusters, the cluster itself contributes statistically to volatility clustering. This is significantly advantageous over other factor models, since the factors can be chosen statistically, whilst also keeping economically relevant factors. Finally, we show that the log volatility factor model explains a similar amount of memory to a Principal Components Analysis (PCA) factor model and an exploratory factor model.    
\end{abstract}

\section{Introduction}
Volatilities are an important factor for the estimation of risk \cite{Bouchaud2009b} and for models aiming at dynamically modelling price and what the rational, fair price should be under such models \cite{Hull1987,Hull2006}. However, the effect of volatility clustering, and particularly its unclear link with how volatilities are correlated with each other, complicates this process. This causes a problem due to the high dimensionality of the correlation matrix between the log volatilities that is also subject to noise \cite{Bun2017}, which makes it difficult to identify meaningful information about what drives the volatility and volatility clustering. This problem is also relevant in multivariate volatility modelling since most popular methods such as multivariate General Autoregressive Conditional Heteroskedasticity (GARCH) \cite{Bauwens2006}, stochastic covariance \cite{Clark1973} and realised covariance \cite{Andersen2003} suffer from the curse of dimensionality and an increase in the number of parameters needed. One such way of tackling this problem is through dimensionality reduction, which is a general class of methods that aims to reduce high dimensional datasets to a reduced form which is a faithful representation of the original dataset \cite{Maaten2009}, and is also related to noise reduction of the dataset.

One such method of dimensionality reduction of correlation matrices is Principal Component Analysis (PCA) \cite{Jolliffe1986}. It aims to transform the original correlation matrix into an orthogonal basis. For square correlation matrices, which are those that we consider in this paper, this essentially means calculating the eigenvalues and their respective eigenvectors. The first eigenvector (called the first principal component) has the highest variance and explains most of the variability in the data, the second eigenvector (called the second principal component) has the second highest variance and explains less variability than the first principal component, and so on. The method has been applied to finance mainly through portfolio optimisation to produce sets of  orthogonal portfolios \cite{Darbyshire2017}. A paper which uses PCA in the context of volatility modelling is \cite{Alexander2002}, where the author extracts the first few principal components and uses them to calibrate a multivariate GARCH model, with a further extension proposed in \cite{Zhang2009}. The main drawback of PCA is that it is not clear how many principal components i.e. factors to keep, as either too many principal components are kept or the methods used to select the components are heuristic and subjective in nature \cite{Jolliffe1986}. In \cite{Plerou2002}, the authors suggest to keep the number of principal components according to the Marchenko-Pastur distribution with a further refinement made in \cite{Majumdar2012} and previously in \cite{Jackson1993}, however in \cite{Livan2011} it is pointed out valuable information may still be lost. 

A highly related class of methods in dimensionality reduction are called factor models \cite{Sharpe1964,Roll1980,Fama1993,Chicheportiche2015}. Factor models are used to describe the dynamical evolution of time series, assuming that there exist common factors through the asset's sensitivity, often called responsiveness, to changes in the value of these factors. Dimensionality reduction is then achieved through the description of the time series as the number of factors is smaller than the number of stocks. Factor models have widespread use in finance due to their relative (or at least superficial) simplicity in comparison to other models of returns series \cite{Sharpe1964,Fama1993,Fama1996,Engel2015,Chicheportiche2015}. Factor models can be split into two varieties: exploratory, which assume no underlying structure to the data, and confirmatory, which tests relationships between known factors \cite{Thompson2004}.

However, similarly to PCA a question of how we should choose the factors arises. One such answer can be categorised by assuming  that we have some prior knowledge of the factors. The simplest and earliest factor model which falls under this category is the Capital Asset Pricing Model (CAPM)\cite{Sharpe1964,Merton1973,Zabarankin2014,Barberis2015}. It emerges from the extremely popular Markowitz scheme of portfolio optimisation \cite{Markowitz1952}, which says it is better to spread an investment across a class of stocks in order to reduce the total risk of the portfolio. CAPM develops this further by saying that the non diversifiable risk, or systematic risk, comes from the stock's exposure to changes in the market and the corresponding sensitivity to this change.

A very well known factor model which has multiple factors, rather than just one like CAPM, is the 3-factor Fama-French factor model \cite{Fama1992,Fama1993,Fama1996,Connor2012,Faff2014}. In this factor model, the first factor comes again from the exposure to the market risk with two extra factors: the small minus big (SMB) and the high minus low (HML)\cite{Fama1993,Fama1996}. The SMB factor follows the observation by Fama and French that stocks with a smaller market cap, which is the market value of the stock used as a proxy of size, tend to give additional returns. Equivalently, the HML factor represents the book/market ratio i.e. the ratio of the total value of the assets owned by the company associated to a stock relative to the stock's market value, and is positively correlated with additional returns. The aim of the HML factor is to evaluate whether stocks have been undervalued by the market, where the book/market ratio exceeds 1, and thus have the potential for larger returns. Recently, the Fama-French model has been extended to include 5 factors \cite{Fama2015}. The arbitrage pricing theory (APT) is also a more generalised multi factor model, except it states that returns are a linear function of macro economic factors \cite{Roll1980,Chen1986}. In APT however, there is no indication of exactly how many and what factors should be included, which then introduces an ad-hoc nature to the types and numbers of factors included in the model.

The above factor models share the fact that the number and nature of the factors are somewhat exogenous in the sense that they are determined by economic intuition on what should drive financial returns. Unfortunately, it has been pointed out that there is weak evidence for CAPM \cite{Fama1992}, both Fama-French $3$ and $5$ factor models and to some manifestations of the APT \cite{Reinganum1981,Faff2004,Grauer2010,Racicot2016}, underlying the issue that these factors cannot explain the cross dependence of assets. Instead, there is a strand of literature which invokes factors that are extracted from the financial data itself thus meaning that the factors are endogenous \cite{Malevergne2004,Tumminello2007,Chicheportiche2015}. In essence, it has been shown that the collective action of assets is what induces the factors, giving support to this type of determination of factors \cite{Malevergne2004}, an approach we shall adopt here. Another difference is that the above factor models are mainly applied to returns rather than volatilities.

In this paper, we instead build a new factor model of log volatilities that aims to reduce the dimensionality by considering contributions globally from the market and more locally to the clusters and their interactions. The number of factors is fixed by the Directed Bubble Hierarchical Tree (DBHT) clustering algorithm \cite{Song2012,Musmeci2015}, which therefore means we make no prior assumption on the number of clusters and thus the number of factors to be considered. Using this factor model between volatilities, we aim to study the link between the univariate volatility clustering and the multivariate correlation structure of volatilities. We will see that whilst over the entire market the only significant contributor that affects the memory is the market, individual clusters may have different properties where the cluster contributions and interactions are more significant. This offers a method to statistically select factors based on memory reduction. We also note that for the clusters which significantly reduce their own memory are mostly made up by stocks from particular industries, offering an economic interpretation for the makeup of the cluster modes. We can thus select the factors in a statistical manner like in PCA, but also retain the appealing economic interpretation like in CAPM and Fama-French.

The structure of the paper is as follows: Section \ref{dataset} desribes the dataset, Section \ref{NewLogVolFactorModel} introduces a new factor model for log volatilities, Section \ref{MiccicheResult} describes how we select factors based on their memory reduction using a new non parametric integrated proxy for the strength of the volatility clustering, Section \ref{MemoryFiltration} we explore how the empirical link between volatility clustering strength and volatility cross correlation can be explained. In Section \ref{EconomicalInterpretation}, we reveal how each cluster has an economical interpretation in terms of their identified dominant ICB supersector. Section \ref{ComparisonPCA} compares our factor model to a PCA inspired factor model and an exploratory factor analysis model in terms of their memory reduction performance. Section \ref{DynamicalStability} reports the dynamic stability of the factor model. Finally, we draw some conclusions in Section \ref{Conclusion}.
\section{Dataset} \label{dataset}
The dataset we shall use consists of the daily closing prices of 1270 stocks in the New York Stock Exchange (NYSE), National Association of Securities Dealers Automated Quotations (NASDAQ) and American Stock Exchange (AMEX) from 01/01/2000 to 12/05/2017, which makes $4635$ points for each price time series. As anticipated in the introduction, we perform cross correlation analysis. We therefore make sure that the stocks are aligned through the data cleaning procedure described in \ref{DataCleaning}, which leaves our dataset with $N=1202$ stocks. We calculate the log-returns time series of a given stock $i$, $r_{i}(t)$, defined as: 
\begin{equation}
r_{i}(t)=\ln p_{i}(t+1) - \ln p_{i}(t),
\end{equation}
where $p_{i}(t)$ is the price time series of stock $i$, and $r_{i}(t)$ is a time series of length $T=4364$. After standardising $r_{i}(t)$ so that it has zero mean and a variance of 1, we define the proxy we shall use for the volatility as $\ln |r_{i}(t)|$ i.e. the log absolute value of returns \cite{Taylor1994}. 

\section{Log-volatility factor model} \label{NewLogVolFactorModel}
In this section we describe a new factor model for log volatilities, which we shall use to uncover the relationship between the univariate volatility clustering effect and the cross correlations between volatilities. Let us recall that a general factor model is given by:
\begin{equation}
r_{i}(t)=\sum_{p=1}^{P}\left[\beta_{ip}f_{p}(t)+\alpha_{ip}\right]+\epsilon_{i}(t), \label{GLM}
\end{equation}
where $r_{i}(t)$ are the log returns for asset $i$, $f_{p}$ are the $p=1,2,...,P$ factors. $\beta_{ip}$ is their respective sensitivities/responsiveness, which quantifies how $r_{i}(t)$ reacts to changes in $f_{p}$. $\alpha_{ip}$ is the intercept and $\epsilon_{i}(t)$ are residual terms with zero mean. Firstly, we define the log volatility term we want to study. Most stochastic volatility models (where the volatility is assumed to be random and not constant) assume that the returns for the stock $i$ follow an evolution according to \cite{Breidt1998}
\begin{equation}
r_{i}(t)=\delta(t)e^{\omega_{i}(t)}, \label{StochVolModel}
\end{equation}
where $\delta(t)$ is a white noise with finite variance and $\omega_{i}(t)$ are the log volatility terms. The exponential term encodes the structure of the volatility and how it contributes to the overall size of the return. Taking the absolute value of \eqref{StochVolModel} and the log of both sides, Eq. \eqref{StochVolModel} becomes
\begin{equation} \label{LogVolTerms}
\ln |r_{i}(t)|=\ln |\delta(t)|+\omega_{i}(t),
\end{equation}
from which we see that working with $\ln |r_{i}(t)|$ has the added benefit of making the proxy for volatility, $\omega_{i}(t)$ additive, which in turn makes the volatility more suitable for factor models. Since $\delta(t)$ is a random scale factor that is applied to all stocks, we can set it to $1$, so that $\omega_{i}(t)=\ln |r_{i}(t)|$. We also standardise the $\ln |r_{i}(t)|$ to a mean of $0$ and standard deviation $1$ as is performed in \cite{Singh2016}. 
In the following subsections, we describe our factor model which considers contributions from the market mode, clusters and interactions, and their corresponding fitting procedures. 

\subsection{Market Mode} \label{MarketMode}
The log volatility term $\omega_{i}(t)$ in Eq. \eqref{LogVolTerms} can be modelled as 
\begin{equation}
\omega_{i}(t)=\beta_{i0}I_{0}(t)+\alpha_{i0}+c_{i}(t), \label{MarketEqn}
\end{equation}
where $\beta_{i0}$ is the responsiveness of stock $i$ with respect to changes in $I_{0}(t)$, defined as
\begin{equation}
I_{0}(t)=\sum_{i=1}^{N}\xi_{i}\ln |r_{i}(t)|, \label{MarketModeWeighted}
\end{equation}
with the pseudo-index $\xi_{i}$ being the weight of stock $i$ for the market mode. $\alpha_{i0}$ in eq. \eqref{MarketEqn} is the excess volatility compared to the market, $I_{0}(t)$. We note that the factor model in eq. \eqref{MarketEqn} is in analogous form to the general factor model in eq. \eqref{GLM}. The first two terms of eq. \eqref{MarketEqn} represent the market factor, which is the widely observed effect of the market affecting all stocks i.e. the co-movement of all stocks at once \cite{Laloux1999,Plerou2002,Singh2016}. We see from eq. \eqref{MarketEqn} that performing the linear regression of $\omega_{i}(t)$ against $I_{0}(t)$ gives $\beta_{i0}$ and $\alpha_{i0}$, so that the $c_{i}(t)$ becomes the residue after performing the regression. In table \ref{table:MarketModeCoeffcients}, we show two examples of the regression coefficients for the market mode for two selected stocks Coca Cola Enterprises (KO) and Transoceanic (RIG). We report the values of $\beta_{i0}$ and $\alpha_{i0}$ for the weighted scheme and for the equal weights scheme detailed in \ref{WeightingSchemes}, along with their p values for the null hypothesis of each of the coefficients being $0$. As we can see from Table \ref{table:MarketModeCoeffcients}, at the 5\% level, the null hypothesis is rejected for all $\beta_{i0}$ for both weighting schemes, which means that we can conclude that the $\beta_{i0}$ are significant. For the $\alpha_{i0}$ the null hypothesis is rejected for both stocks in the equal weights case, and for the weighted case it is rejected only for RIG, and for these cases we can conlude that the $\alpha_{i0}$ are non-zero.

\begin{table}[htbp]
\begin{subtable}{0.45\linewidth}
\centering
{\begin{tabular}{|l|l|l|}
    \hline
          & $\beta_{i0}$   & $\alpha_{i0}$ \\
    \hline
    KO    & 0.0310 (0) & 0.0015 (0.4764) \\
    \hline
    RIG   & 0.0248 (0) & 0.1972 (0) \\
    \hline
    \end{tabular}}
\caption{weighted modes}
\label{table:WeightedMarket}
\end{subtable}
\hfill
\begin{subtable}{0.45\linewidth}
\centering
{\begin{tabular}{|l|l|l|}
    \hline
          & $\beta_{i0}$   & $\alpha_{i0}$ \\
    \hline
    KO    & 1.1564 (0) & -0.0690 (0.0017) \\
    \hline
    RIG   & 0.9041 (0) & 0.1426 (0) \\
    \hline
    \end{tabular}}
\caption{equal weights}
\label{table:EqualWeightsMarket}
\end{subtable}
\caption{This table shows the responsiveness to the market mode $I_{0}(t)$, $\beta_{i0}$ and the corresponding excess volatility $\alpha_{i0}$ for stocks KO and RIG, calibrated as detailed in section \ref{MarketMode}. The p values shown in brackets are for the null hypothesis that both $\beta_{i0}$ and $\alpha_{i0}$ are 0. Table \ref{table:WeightedMarket} is for the weighted scheme and Table \ref{table:EqualWeightsMarket} for equal weights, which are detailed in \ref{WeightingSchemes}.
\label{table:MarketModeCoeffcients}}
\end{table} 
 
\subsection{DBHT output} \label{DBHTOutput}
Since $c_{i}(t)$ is the residue after performing the regression in eq. \eqref{MarketEqn}, it represents the volatility that is not explained by the market. We can therefore further define $c_{i}(t)$ as:
\begin{equation}
c_{i}(t)=\beta_{ik}I_{k}(t)+\sum_{k'=1}^{n-1}\beta_{ik'}I_{k'}(t)+\epsilon_{i}(t), \label{Residue}
\end{equation}
where $\beta_{ik}$ are the responsiveness for the $k$ cluster mode $I_{k}(t)$ which $i$ is a member of. In the sum from eq. \eqref{Residue}, the $\beta_{ik'}$ are the responsiveness to changes in $I_{k'}(t)$ which are the cluster modes of the clusters $k'\neq k$ i.e. the clusters $i$ is not a member of. In eq. \eqref{Residue} the first term is for the cluster factor and it represents the co-movement of the stock with its cluster. Like for eq. \eqref{MarketEqn}, eq. \eqref{Residue} is an analogous form to eq. \eqref{GLM}. The sum in eq. \eqref{Residue} represents the interactions the stock $i$ has with other clusters, where the strength of the interactions are quantified and defined through the $\beta_{ik'}$.

The next step of the calibration procedure concerns the identification of the clusters, which is relevant for the $c_{i}(t)$ term defined in eq. \eqref{Residue}.  Now, we need to find what the cluster structure is, which we do by first calculating $\mathbf{G}$, which is the cross correlation matrix between $c_{i}(t)$, defined as 
\begin{equation}
G_{ij}=\frac{1}{T}\sum_{t=1}^{T}c_{i}(t)c_{j}(t). \label{CorrResidue}
\end{equation}
We then apply the clustering algorithm to $\mathbf{G}$. We use the clustering algorithm after removing the market mode since this gives a more stable clustering \cite{Borghesi2007}. We shall use the Directed Bubble Hierarchical Tree, DBHT \cite{Song2012,Musmeci2015,Musmeci2016}, to find the cluster membership of stocks. DBHT is used because as compared to other hierarchical clustering algorithms it provides the best performance in terms of information retrieval \cite{Musmeci2015}. Using the DBHT algorithm also means that we make no prior assumption on exactly how many factors for the clusters should be included, instead extracting them directly from the data. We can see from Table \ref{table:ClusterSectorSig} that the DBHT algorithm identifies a total of $K=29$ clusters, with the largest cluster comprising of $172$ stocks and the smallest cluster comprising of $5$ stocks. The average cluster size is $41.4$. 
\subsection{Cluster modes and interactions} \label{ClusterModeCal}
Once the number and composition of each cluster is identified, we can associate a factor to each cluster $k$. The interactions are then characterised through the responsiveness $\beta_{ik'}$ where $k\neq k'$ i.e. how $c_{i}(t)$ changes w.r.t to $I_{k'}(t)$. We define the cluster mode for cluster $k$, $I_{k}(t)$, again as a weighted average of volatilities for the assets in $k$ 
\begin{equation}
I_{k}(t)=\sum_{i\in cluster \ k}\xi_{ik}c_{i}(t). \label{ClusterModeWeighted}
\end{equation}
$\xi_{ik}$ is the weight for stock $i$ which is in cluster $k$. From eq. \eqref{Residue}, we see that similarly to the market mode case, we can determine $\beta_{ik}$, $\beta_{ik'}$ and $\alpha_{ik}$, $\alpha_{ik'}$ by linearly regressing $c_{i}(t)$ against $I_{k}(t)$ and $I_{k'}(t)$. We use elastic net regression \cite{Zou2005} to find $\beta_{ik}$ and $\beta_{ik'}$ to take into account the possibility of $I_{k}(t)$ and $I_{k'}(t)$ being correlated, whilst also allowing for some of the $\beta_{ik'}$ to be $0$ as $i$ may not interact with cluster $k'$. More details about elastic net regression are provided in appendix \ref{ElasticNet}.  
   
\section{Empirical link between volatility clustering and volatility cross correlation} \label{MiccicheResult}
As anticipated in the introduction, we choose which factors are relevant for the decomposition in Eq. \eqref{Residue}, by measuring what the impact is of each cluster on the volatility clustering. Before turning our attention to this analysis, let us introduce the volatility clustering proxy we use in the rest of the paper.
\subsection{Volatility Clustering}
\begin{figure}
\centering
\begin{subfigure}{0.475\textwidth}
	\centering
	\includegraphics[width=\textwidth]{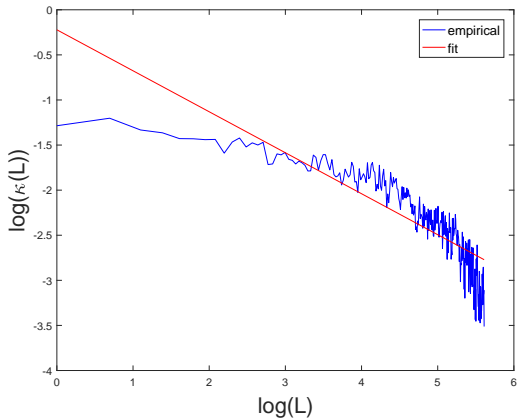}
	\caption{Coca Cola Enterprises Inc. $\beta^{\text{vol}}=0.4544$}
	\label{fig:ACF_KO}
\end{subfigure}
\hfill
\begin{subfigure}{0.475\textwidth}
	\centering
	\includegraphics[width=\textwidth]{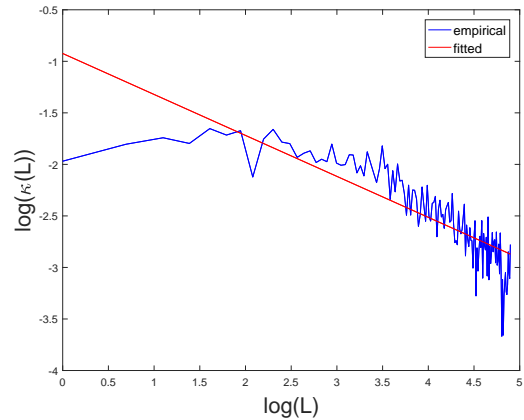}
	\caption{Transoceanic $\beta^{\text{vol}}=0.3975$}
	\label{fig:ACF_RIG}
\end{subfigure}
\caption{Empirical ACF of the log absolute value returns (blue solid lines) for Coca Cola Co. (KO) in figure \ref{fig:ACF_KO} and Transocean (RIG) in figure \ref{fig:ACF_RIG} in log-log scale. The linear best fit is also shown in red dashed lines. }
\label{fig:ACFExamples}
\end{figure}

Volatility clustering is one of the so called stylised facts of financial data, and expresses the idea that returns are not independent since volatilities are autocorrelated \cite{Cont2001,Chakraborti2011}. The autocorrelation function (ACF) $\kappa(L)$ is defined as
\begin{align}
\kappa(L)&=corr(\ln |r(t+L)| ,\ln |r(t)|) \\
&=\frac{\langle\left[\ln |r(t+L)| \ln |r(t)|\right]\rangle}{\sigma^{2}},
\end{align}
where $\langle...\rangle$ denotes the expectation. $L$ is the lag and $\sigma^{2}$ is the variance of the process of $\ln |r(t)|$, and note that we use log absolute value returns as a proxy for volatility. The interpretation of this result is that large changes in returns are usually followed by other large changes in returns, or that the returns retain a memory of previous values \cite{Mandelbrot1997}. For this reason, volatility clustering can also be called the memory effect. $\kappa(L)$ has also been assumed to follow a power law decay:
\begin{equation}
\kappa(L)\sim L^{-\beta^{\text{vol}}}, \label{PowerLawACF}
\end{equation}
where $\beta^{\text{vol}}$ describes the strength of the memory effect. A lower value of $\beta^{\text{vol}}$ indicates that more memory of past values is kept. To compute $\beta$ we transform eq. \eqref{PowerLawACF} into loglog scales and compute the slope of the linear best fit, which gives us the exponent $\beta^{\text{vol}}$. We shall compute $\beta^{\text{vol}}$ using the Theil-Sen procedure rather than using standard least squares since it is more robust to outliers \cite{Theil1992}. We report in figure \ref{fig:ACFExamples} the function $\kappa(L)$ for Coca Cola Enterprises Inc. in figure \ref{fig:ACF_KO} and Transoceanic in figure \ref{fig:ACF_RIG}, both in loglog scale, with the linear best fit also plotted.  
We define the entries $E_{ij}$ of the empirical volatility cross correlation $\mathbf{E}$ as
\begin{equation}
E_{ij}=\sum_{t=1}^{T}\ln |r_{i}(t)|\ln |r_{j}(t)| . 
\end{equation}  
The proxy used for the volatility cross correlation is the average cross correlation for stock $i$, $\rho_{i}^{\text{vol}}$, is defined as
\begin{equation}
\rho_{i}^{\text{vol}}=\frac{1}{N-1}\sum_{i\neq j}^{N}E_{ij}
\end{equation} 
\begin{figure}
\centering
\includegraphics[width=0.75\textwidth]{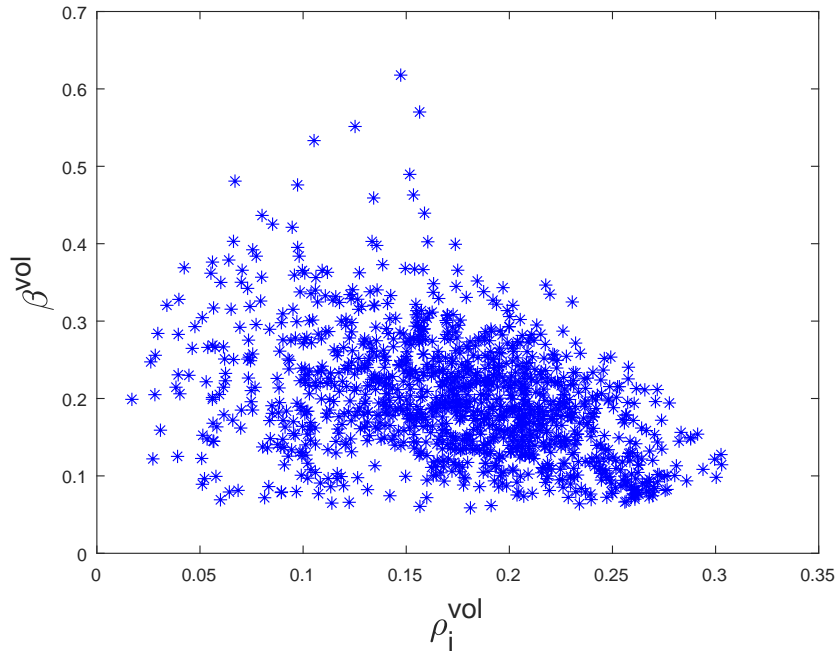}
\caption{Negative dependence between $\rho_{i}^{\text{vol}}$ and $\beta_{i}^{\text{vol}}$. The negative relationship was tested using 1 sided Spearman's rank correlation at the 5\% level with the null hypothesis of there being no correlation and was rejected, which confirms the result of \cite{Micciche2013} on our data.}
\label{fig:MiccicheResult}
\end{figure}
Using the proxies for volatility clustering and the volatility cross correlation, \cite{Micciche2013} finds a negative relationship between $\rho_{i}^{\text{vol}}$ and $\beta_{i}^{\text{vol}}$, which we confirm holds on our data set of daily data and using $\ln |r(t)|$, rather than the original high frequency data and $|r(t)|$ used in \cite{Micciche2013}, in figure \ref{fig:MiccicheResult}. The main consequence of this result is that it implies that the more the volatility of a stock $i$ is linked to other stocks, the stronger the memory effect and thus it retains more information about previous values of volatility, linking the strength of volatility clustering with the cross correlation matrix between volatilities.  

\subsection{Non parametric memory proxy} \label{IntegratedProxy}
\begin{figure}
\centering
\begin{subfigure}{0.475\textwidth}
\includegraphics[width=\textwidth]{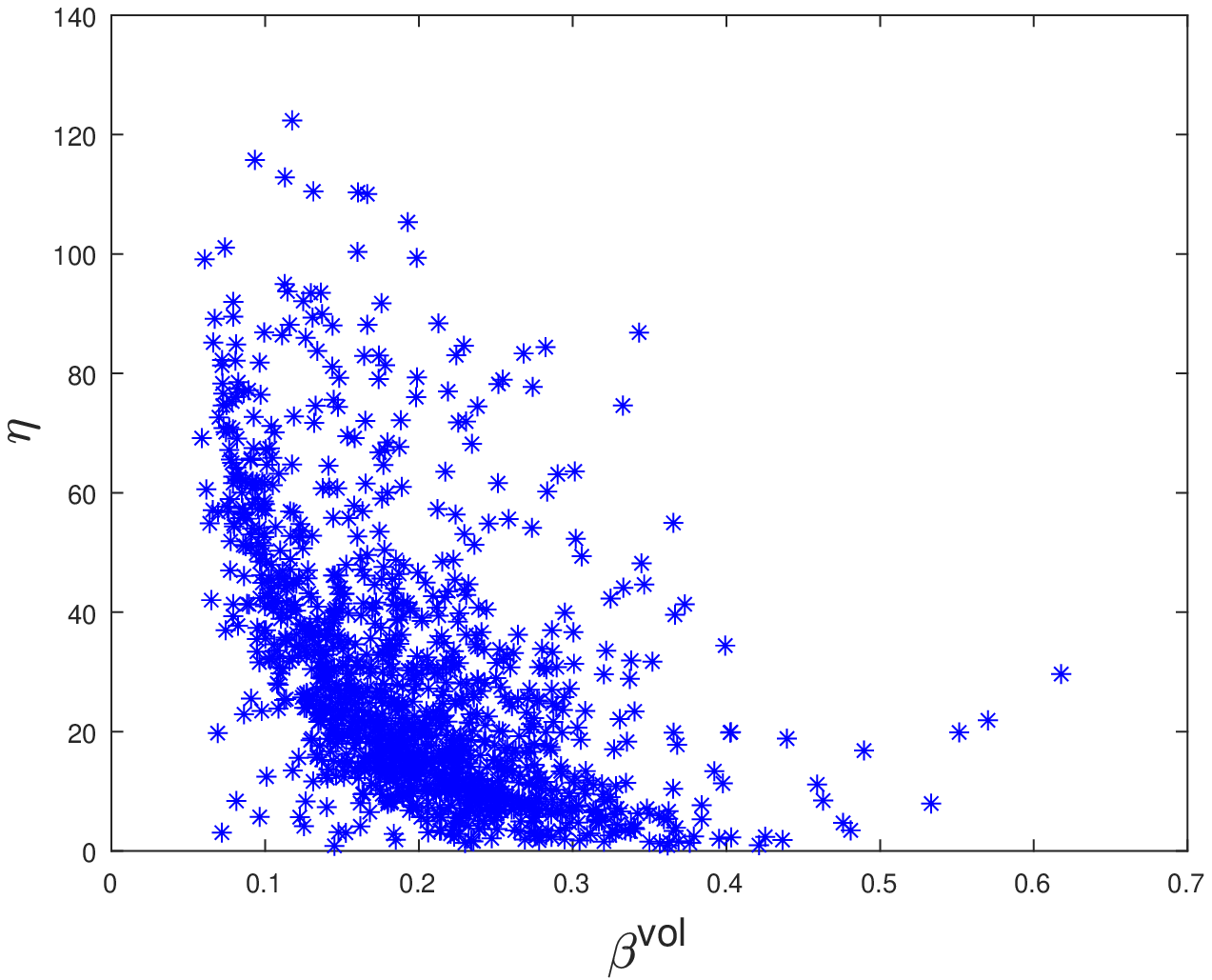}
\caption{$\beta^{\text{vol}}$ vs $\eta$}
\label{fig:IntProxy_BetaProxy}
\end{subfigure}
\hfill
\begin{subfigure}{0.475\textwidth}
\includegraphics[width=\textwidth]{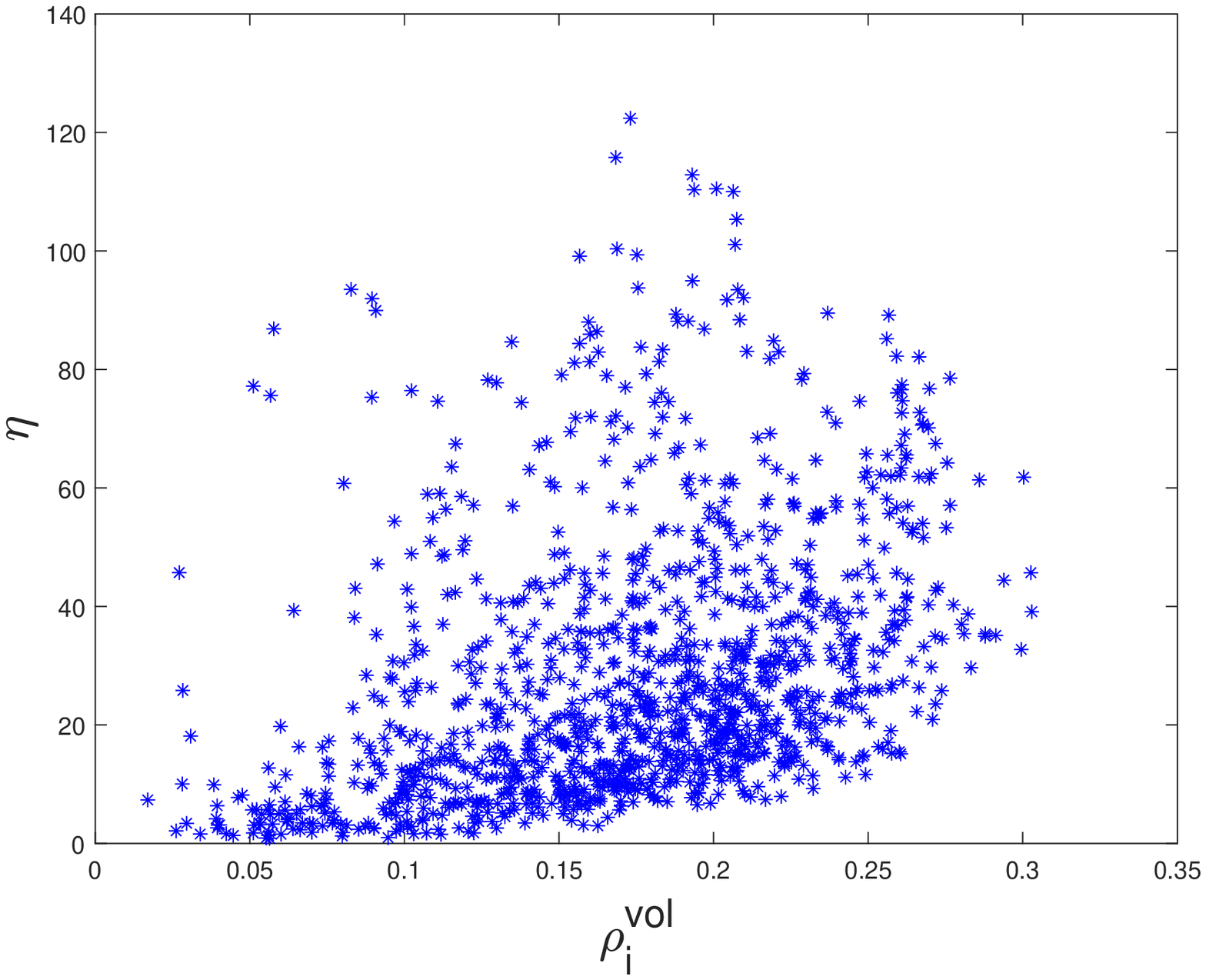}
\caption{$\rho_{i}^{\text{vol}}$ vs $\eta$}
\label{fig:IntProxy_AvgCorr}
\end{subfigure}
\caption{In figure \ref{fig:IntProxy_BetaProxy} we plot the $\beta^{\text{vol}}$ power law exponent proxy for the strength of the memory effect vs c the integrated proxy. In figure \ref{fig:IntProxy_AvgCorr} we plot the relationship between $\rho_{i}^{\text{vol}}$ and $\eta$ defined in the text. The decreasing relationship in figure \ref{fig:IntProxy_BetaProxy} and the increasing relationship in figure \ref{fig:IntProxy_AvgCorr} was tested using the Spearman's rank correlation at the 5\% level and was rejected in both cases.}
\label{fig:IntProxyProof}
\end{figure}

As already mentioned, the $\beta^{\text{vol}}$ power law exponent that is fitted to the autocorrelation function of the absolute returns is a proxy for the strength of the memory effect: the lower the beta the stronger the memory effect. The use of the power law to quantify the memory effect is parametric as we \emph{assume} the tail decays as a power law through the exponent $\beta$. The autocorrelation function itself can be noisy due to its slow convergence \cite{Cont2001}, which can be seen in figure \ref{fig:ACFExamples}. In light of this, we instead introduce a new model free proxy, $\eta$, by integrating the autocorrelation function over time lags $L$ until $L_{cut}$, which we define as the standard Bartlett Cut at the 5\% level \cite{Box2015}. 
\begin{equation}
\eta=\int_{L=1}^{L_{cut}}\kappa(L)dL \ ,\label{IntProxy}
\end{equation} 
where $\kappa(L)$ is the empirical autocorrelation matrix of the log absolute returns as a function of the lag $L$. With this proxy the larger the value of $\eta$ the greater the degree of the memory effect (in the $\beta$ proxy this corresponds to larger values of the exponent). The median value reported across all stocks is $20.7318\pm 8.6901$, where the error is computed across all stocks using the median absolute deviation (MAD) for $\eta_{i}$ defined as
\begin{equation}
MAD=median\left(\left|\eta_{i}-median(\eta_{i})\right|\right). 
\end{equation}
We have also plotted the $\beta$ as a memory effect proxy vs $\eta$ in figure \ref{fig:IntProxy_BetaProxy}, which as expected shows a decreasing relationship between $\eta$ and the $\beta$ memory proxy, which is the one used in the literature, since a larger memory effect means a higher $\eta$, but lower $\beta$. This provides justification for our use of $\eta$. This proves that $\eta$ is coherent with $\beta^{\text{vol}}$ and thus can be used a proxy for the strength of the memory effect. 

\begin{figure}
\centering
\begin{subfigure}{0.475\textwidth}
\includegraphics[width=\textwidth]{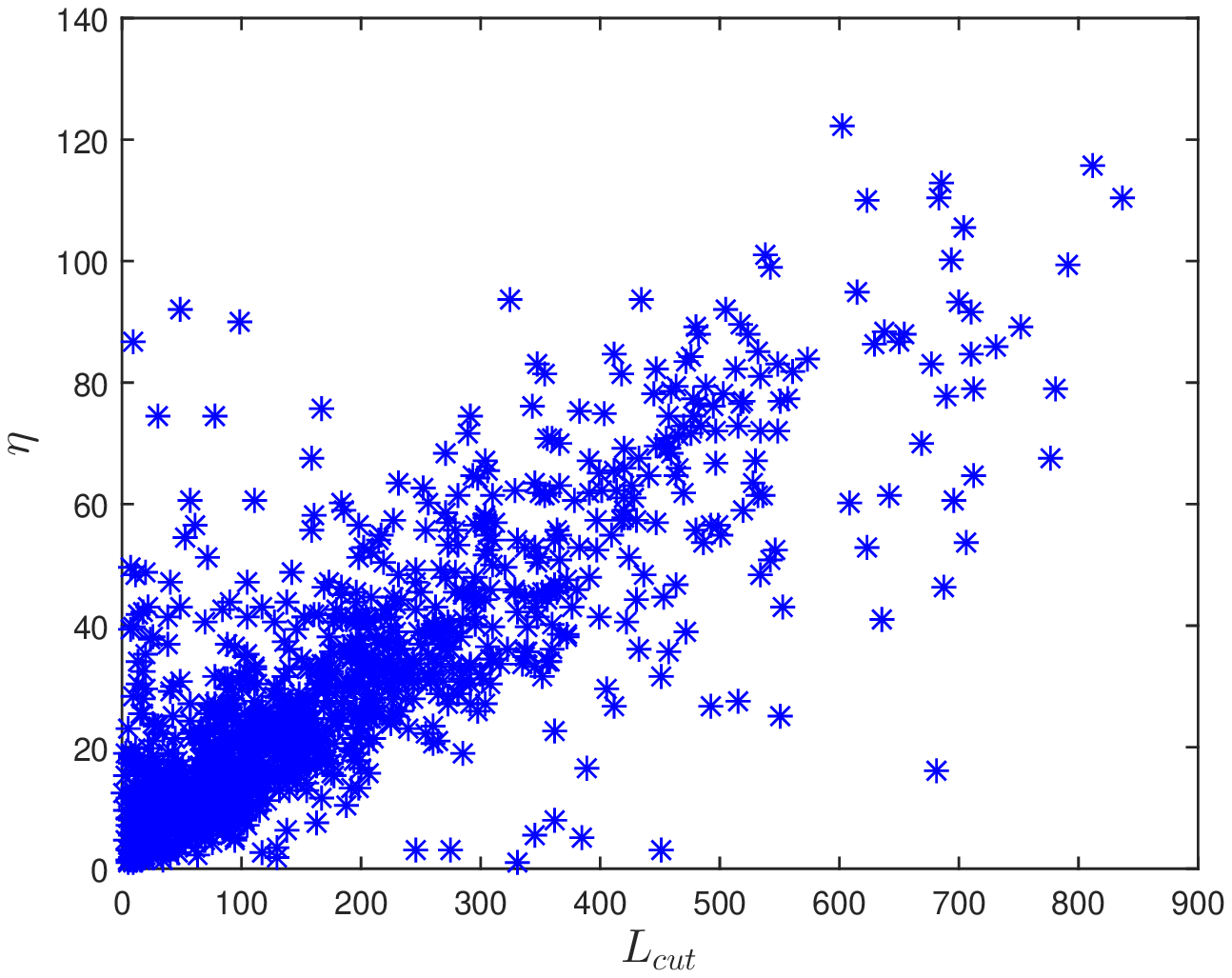}
\caption{$L_{cut}$ vs $\eta$}
\label{fig:LCut_IntProxy}
\end{subfigure}
\hfill
\begin{subfigure}{0.475\textwidth}
\includegraphics[width=\textwidth]{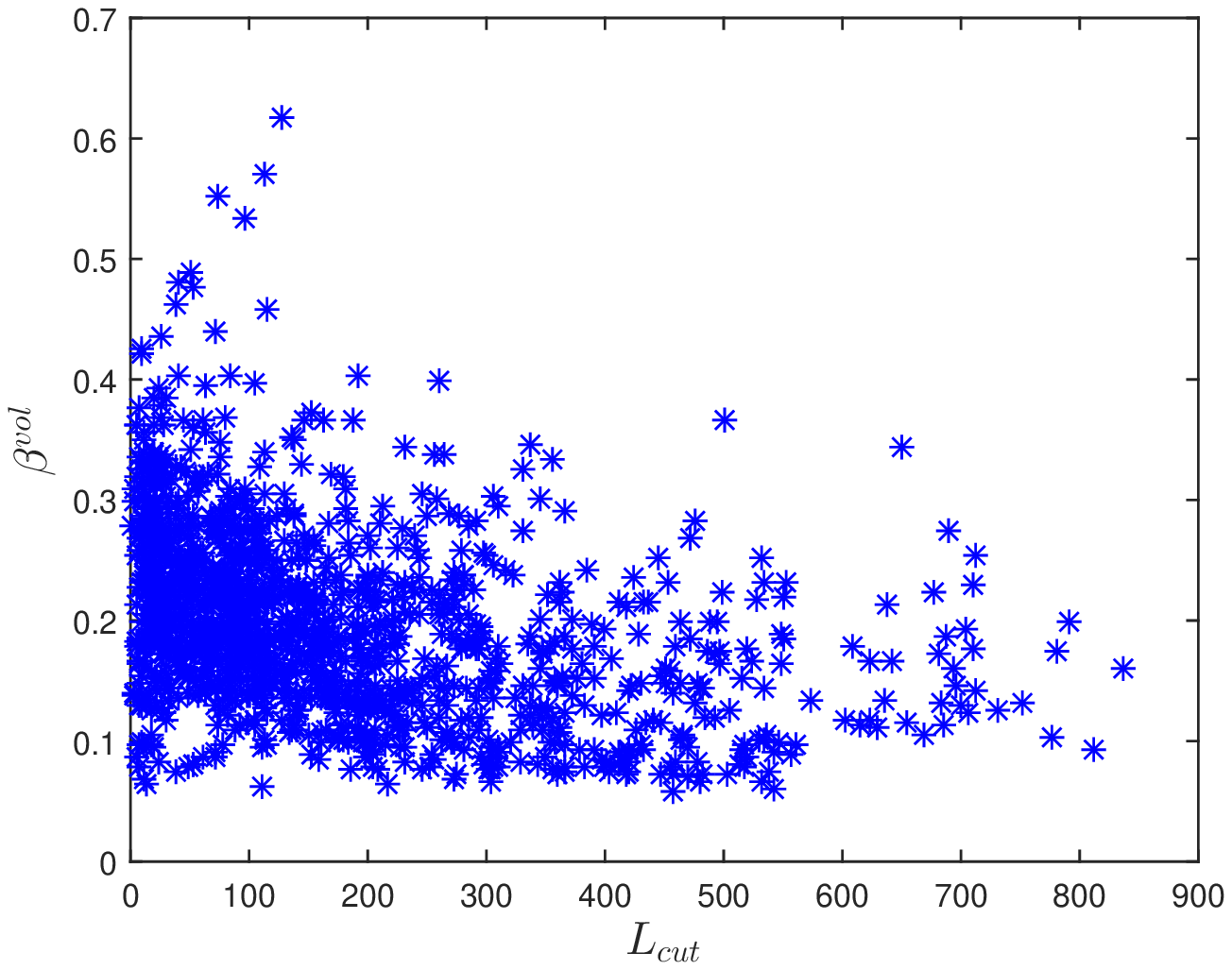}
\caption{$L_{cut}$ vs $\beta^{\text{vol}}$}
\label{fig:LCut_BetaProxy}
\end{subfigure}
\caption{The figure on the left is a plot of $L_{cut}$ vs $\eta$ for all stocks. The figure on the right is $L_{cut}$ vs $\beta^{\text{vol}}$ for all stocks. The increasing relationship shown in figure \ref{fig:LCut_IntProxy} and decreasing relationship shown in figure \ref{fig:LCut_BetaProxy} are tested using Spearman's rank correlation, and are 0.7871 and -0.4271 respectively, which are statistically significant at the 5\% level. }
\label{fig:LCut_Proxies}
\end{figure}  
Figure \ref{fig:IntProxy_AvgCorr} which is a plot of $\rho_{i}^{vol}$ vs $\eta$ confirms the main result of \cite{Micciche2013} using $\eta$ instead of $\beta^{\text{vol}}$, and was tested using Spearman's rank correlation at the 5\% level for the null hypothesis, which was rejected, of there being no correlation between $\rho_{i}^{vol}$ and $\eta$ versus alternative hypothesis of there being significant positive relationship. Our proxy can therefore also confirm the result of \cite{Micciche2013}.

Plotting $L_{cut}$ vs $\eta$ in figure \ref{fig:LCut_IntProxy}, reveals that processes with strong short memory will have a lower $L_{cut}$ and thus lower $\eta$, whilst processes with a long memory component will have higher $L_{cut}$ and $\eta$. This is important since volatility clustering is a result of long memory present in time series. An analogous plot of $L_{cut}$ vs $\beta^{\text{vol}}$ in figure \ref{fig:LCut_BetaProxy} shows the expected decrease in $\beta^{\text{vol}}$ as $L_{cut}$ increases, but the relationship is not as strong as that of $L_{cut}$ vs $\eta$ (an absolute Spearman correlation value of 0.4271 vs 0.7871 tested at the 5\% level). A consequence of this is that $\eta$ can better distinguish between short and long memory processes as compared to $\beta^{\text{vol}}$. 

\section{Memory filtration} \label{MemoryFiltration}
In this section, by means of the factor model introduced in eqs. \eqref{MarketEqn}\eqref{Residue} and also by means of the $\eta$ proxy introduced in the previous subsection, we want to understand the origin of the empirical link between the memory strength and the volatility cross-correlation. This analysis will in turn be also fundamental for the cluster mode selection in our model. The main intuition is that the market mode, the cluster mode and the interaction modes all bring relevant information about the memory of a certain stock's time-series.

\subsection{Assessing the memory contributions}\label{criterion}
Let us here describe the method we use in order to understand the contribution to the memory of each term in the factor model in eqs. \eqref{MarketEqn}\eqref{Residue}. For every time-series, say for stock $i$, we follow a step-by-step procedure, by measuring the value of the proxy $\eta_i$ for the following four times:
\begin{enumerate}
\item on the plain time-series $\eta_{i,PL}$;
\item on the residual time-series once the market mode is removed $\eta_{i,MM}$;
\item on the residual time-series once the market mode and the cluster mode (of the the cluster the stock belongs to) are removed $\eta_{i,CM}$;
\item on the residual time-series once market, cluster and interaction mode are all removed. In order to make a quantitative comparison $\eta_{i,IM}$.
\end{enumerate}
The next step consists in assessing the memory reduction after each removal. We do so by computing the ratio of two subsequently computed value of $\eta_i$. For stock $i$ thus we have that
\begin{enumerate}
\item $\frac{\eta_{i,MM}}{\eta_{i,PL}}$ defines the reduction in memory induced by the market mode;
\item $\frac{\eta_{i,CM}}{\eta_{i,MM}}$ defines the reduction in memory induced by the cluster mode once the market mode is removed;
\item $\frac{\eta_{i,IM}}{\eta_{i,CM}}$ defines the reduction in memory induced by the interaction mode once the market mode and the cluster mode are removed.
\end{enumerate}
According to the definition, if a ratio is below one it means that a memory reduction has occurred via the corresponding removal. In order to understand what is the average behaviour of these ratios we take the median of each of them computed on all stocks. So, for example, the average reduction of memory induced by the market mode on a given set of stocks is $median(\frac{\eta_{i,MM}}{\eta_{i,PL}})$ computed over the index $i$. As for an error to associate to this measure we used the Median Average Deviation \cite{Sachs2012}, defined as for $\frac{\eta_{i,MM}}{\eta_{i,PL}}$
\begin{align} 
&MAD\left(\frac{\eta_{i,MM}}{\eta_{i,PL}}\right) \\
&=median\left(\left|\frac{\eta_{i,MM}}{\eta_{i,PL}}-median\left(\frac{\eta_{i,MM}}{\eta_{i,PL}}\right)\right|\right) \label{MADErrorBar}\ ,
\end{align}
and similarly for $\frac{\eta_{i,CM}}{\eta_{i,MM}}$ and $\frac{\eta_{i,IM}}{\eta_{i,CM}}$. Both the median and the MAD were chosen because of their robustness against outliers. We regard as significant a reduction of memory on the given set of stocks for which the median plus the mad of the ratio are below one.

\subsection{Whole market analysis: finding the main source of memory}
\begin{figure}
\begin{subfigure}{0.475\textwidth}
\centering
\includegraphics[width=\textwidth]{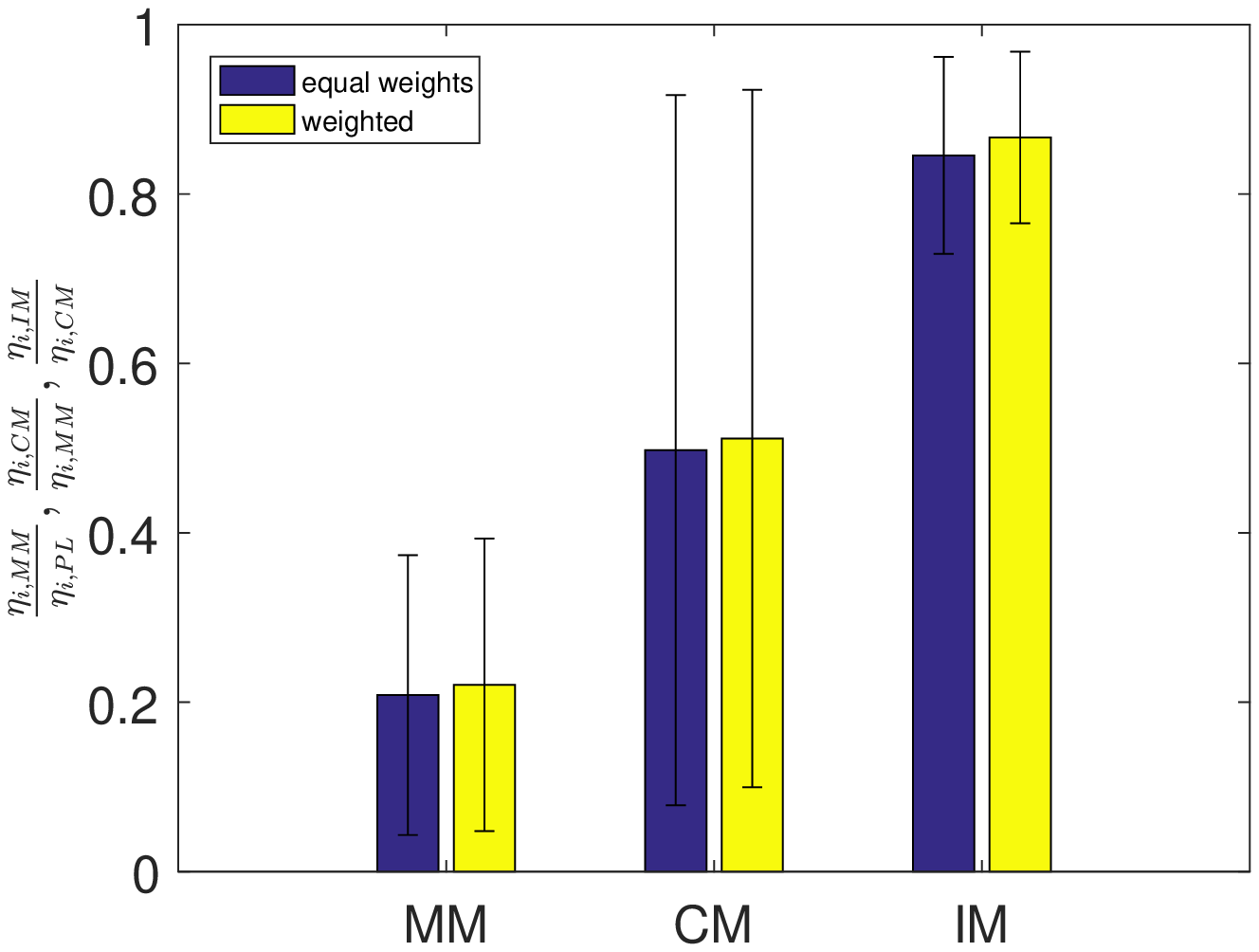}
\caption{\label{fig:medianvolproxy}}
\end{subfigure}
\begin{subfigure}{0.475\textwidth}
\centering
\includegraphics[width=\textwidth]{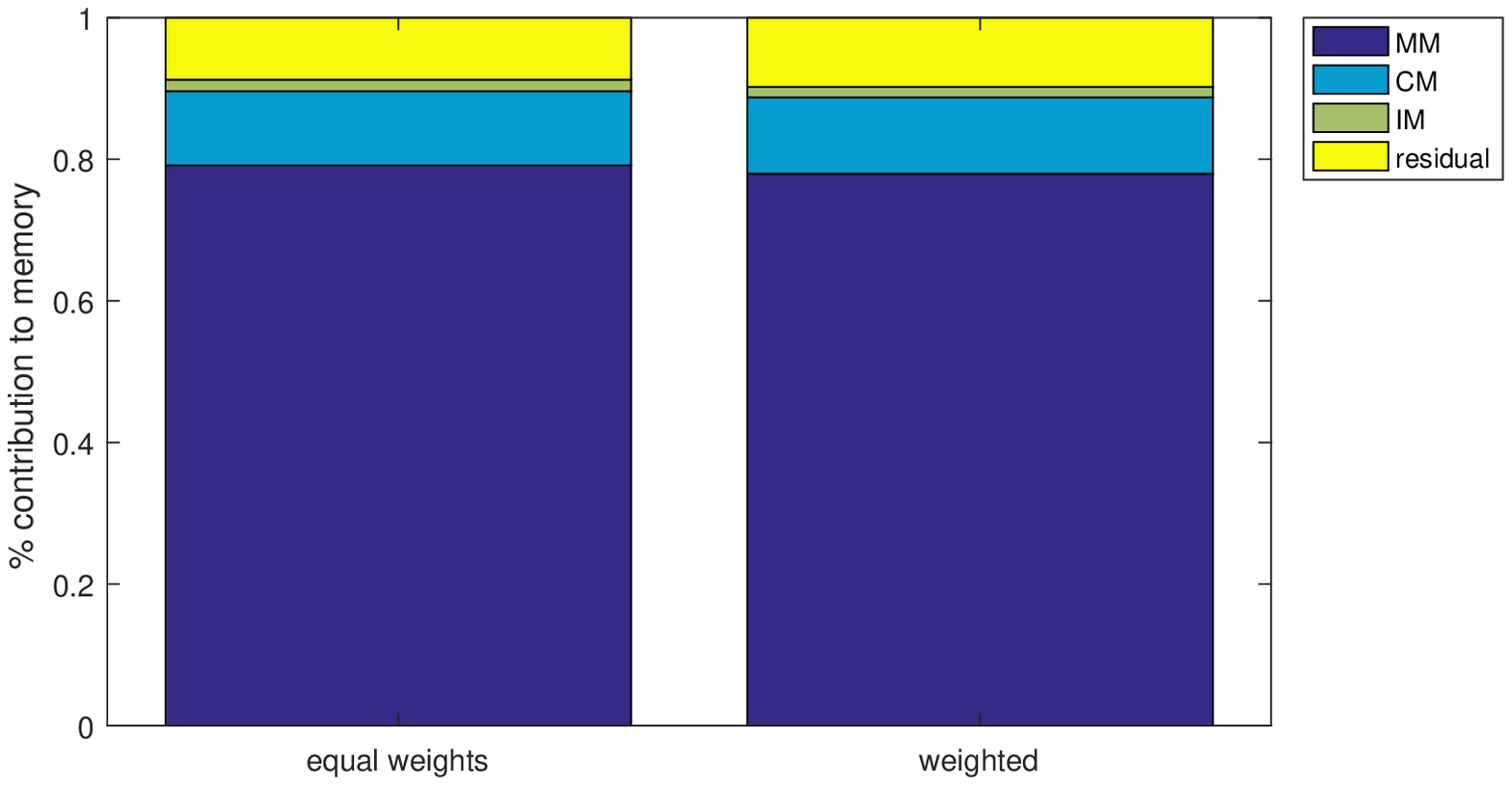}
\caption{\label{fig:contributionmarket}}
\end{subfigure}
\caption{Results for the procedure described in section \ref{criterion} across all stocks in the market. Figure \ref{fig:medianvolproxy} is the median of the ratio of the memory proxies for, starting from the left, $\frac{\eta_{i,MM}}{\eta_{i,PL}}$, $\frac{\eta_{i,CM}}{\eta_{i,MM}}$ and $\frac{\eta_{i,IM}}{\eta_{i,CM}}$, computed over the whole market. The blue bars are for the equal weights scheme and the yellow bars are for the weighted scheme. The black vertical bars represent the errors among stocks memory reduction applied to the whole market, which is calculated using eq. \eqref{MADErrorBar} and its equivalents for the other ratios. In figure \ref{fig:contributionmarket} we plot the contribution to the memory effect of the market (MM), cluster (CM) and interactions (IM) as a percentage with respect to the overall memory. The residual is remaining percentage of memory that is unexplained by the contributors. The values are computed over the whole market. The left column is for the equal weights scheme and the right column is for the weighted scheme.}
\label{fig:marketanalysis}
\end{figure}

\begin{figure}
\centering
\includegraphics[width=0.75\textwidth]{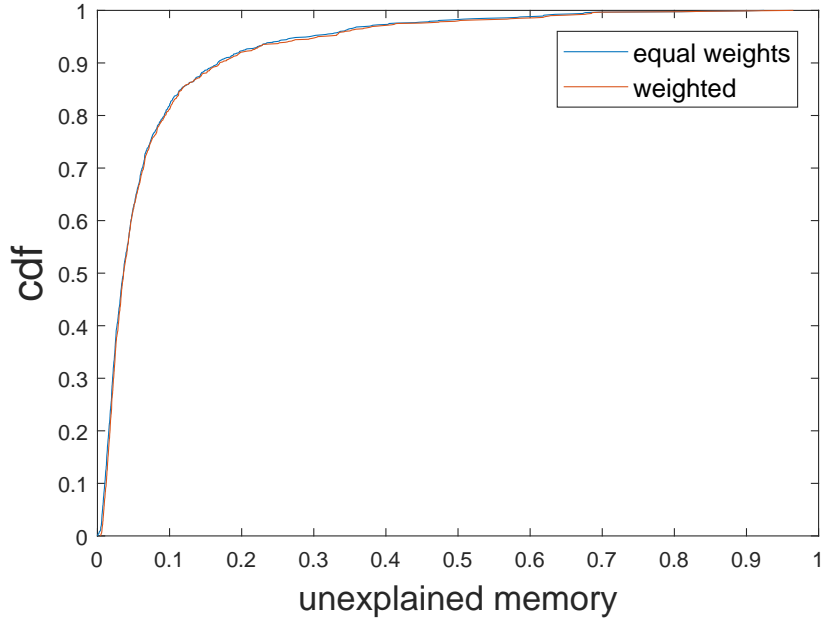}
\caption{Cumulative distribution of the fraction of stocks which have a fraction residual memory left after all contributors of the model (market mode, cluster mode and interactions) are removed. The red line is for the weighted modes and the blue the equal weighted modes}
\label{fig:cumintproxy}
\end{figure}

We apply here the procedure described in the previous subsection to our dataset described in Section \ref{dataset}. For completeness, in Fig. \ref{fig:marketanalysis} we report the result of our analysis for both the unweighted and the  weighted schemes. Figure \ref{fig:medianvolproxy} reports the value of the ratios along with the errors (black vertical bars). We observe that in all cases the average value plus the error stays below one, which means that every term gives a meaningful contribution to the overall memory. However we also notice that, in particular for the reduction coming from the cluster mode, there is a large variablity among stocks. Figure \ref{fig:contributionmarket} reports the same result but showing what is the contribution of each removal with respect to the overall memory. According to our analysis, the majority of the contribution comes from the market mode, which is than the main source of memory for the volatility. We also plot in figure \ref{fig:cumintproxy} the cumulative of the fraction of stocks with at most the percentage of memory left reported on the x axis, after all contributions are removed. For example from figure \ref{fig:cumintproxy} we find that 90\% of all stocks have only 16.7\% of their memory unexplained by all the contributions. We also note here that there is little difference in figure \ref{fig:cumintproxy} between the weighted and unweighted versions so we shall herein use the unweighted scheme for most of the analysis. This analysis establishes that there is indeed a link between the log volatility and volatility clustering.


\subsection{Cluster-by-cluster analysis: selection criterion for factors} \label{ClusterByCluster}
\begin{figure}
\centering 
\begin{subfigure}{0.475\textwidth}
\centering
\includegraphics[width=\textwidth]{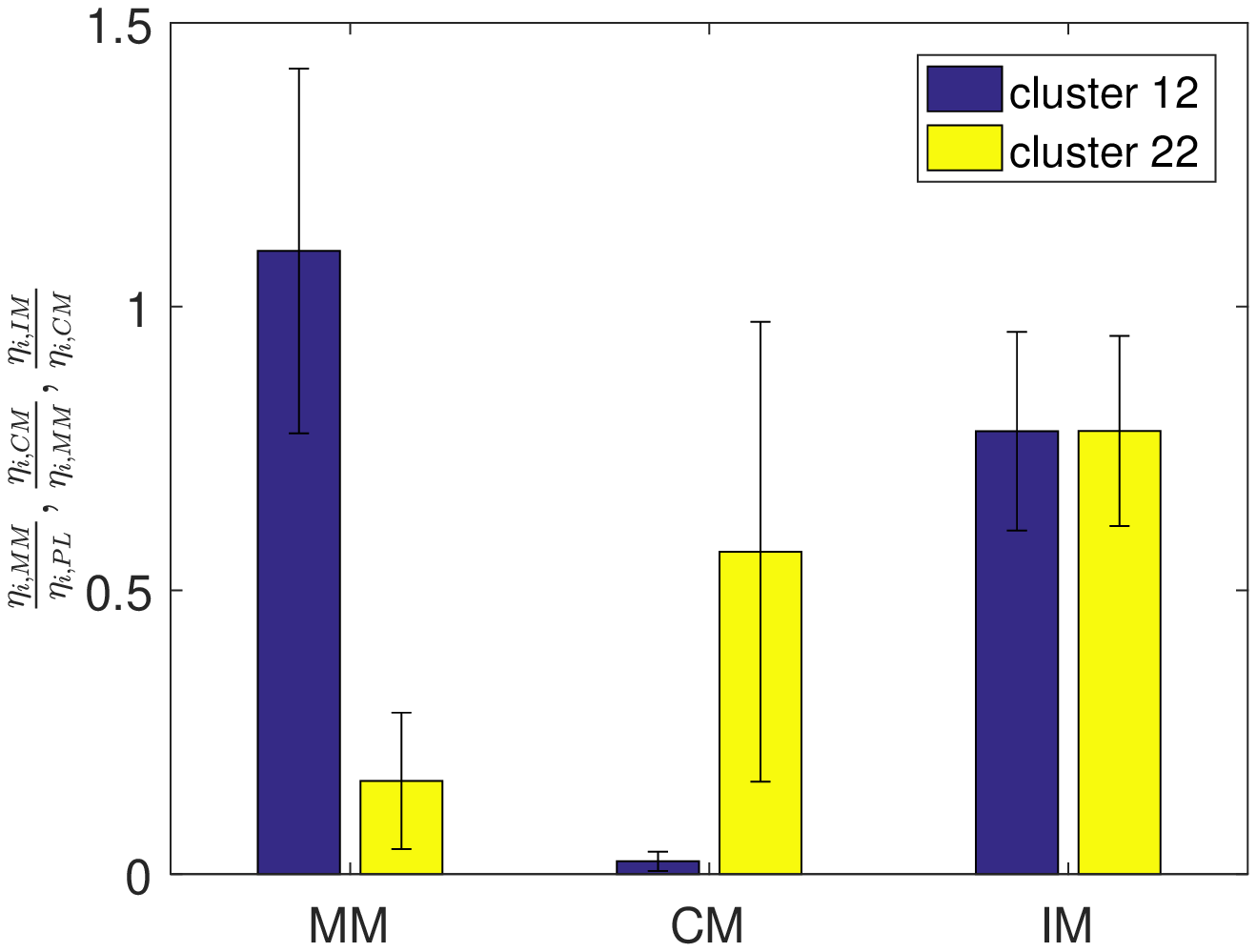}
\caption{\label{fig:medianproxy_cluster12_22}}
\end{subfigure}
\hfill
\begin{subfigure}{0.475\textwidth}
\centering
\includegraphics[width=\textwidth]{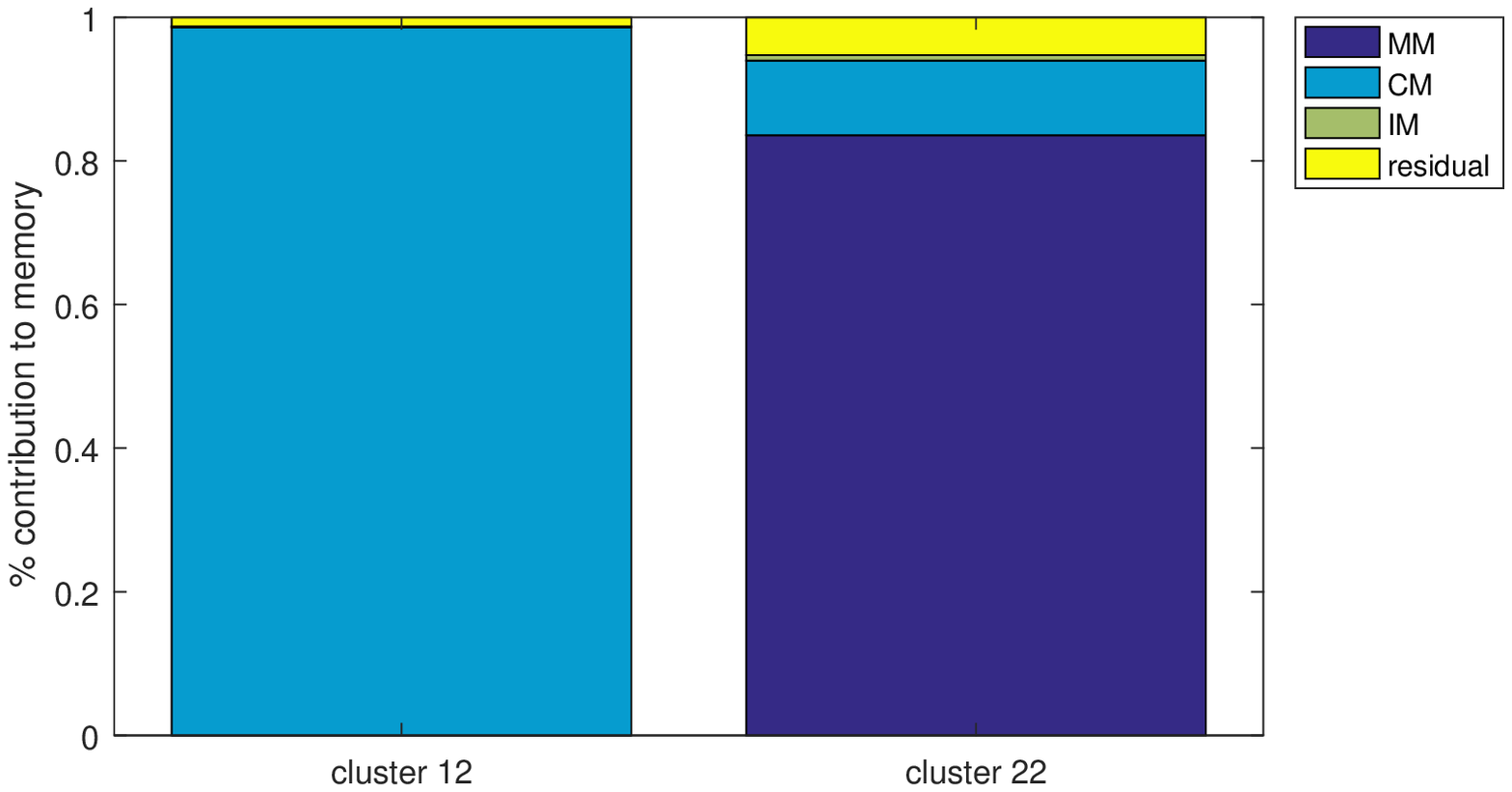}
\caption{
\label{fig:ContributionMemory_cluster12_22}}
\end{subfigure}
\caption{The same set of graphs as Fig. \ref{fig:marketanalysis} except using the equal weights scheme and taking only stocks belonging to cluster 12 and 22. In figure \ref{fig:medianproxy_cluster12_22} we plot the median ratio of, starting from the left, $\frac{\eta_{i,MM}}{\eta_{i,PL}}$, $\frac{\eta_{i,CM}}{\eta_{i,MM}}$ and $\frac{\eta_{i,IM}}{\eta_{i,CM}}$, computed over the stocks in cluster 12 for the blue bars and over stocks in cluster 22 for the yellow bars. Equal weighted modes are used.  The black vertical bars represent the errors among stocks memory reduction applied to stocks in cluster 12 and 22, which is calculated by using eq. \eqref{MADErrorBar} and its equivalents for the other ratios. In figure \ref{fig:ContributionMemory_cluster12_22} we plot the contribution to the memory effect of the market (MM), cluster (CM) and interactions (IM) as a percentage with respect to the overall memory. The residual is remaining percentage of memory that is unexplained by the contributors. The values are computed over all stocks in cluster 12 for the left column and over all stocks in cluster 22 for the right column. Equal weighted modes are used.}
\label{fig:clusterbyclusteranalysis}
\end{figure}    
In this subsection, instead of aggregating the result of the memory reduction over the whole market, we specialize and check what happens to the memory on a cluster-by-cluster basis. For brevity, we only discuss in detail the case of cluster 12 and cluster 22, as defined by the DBHT algorithm discussed in section \ref{DBHTOutput}, since they are quite informative about the different behaviour one can find at a cluster level. We repeat then the same analysis we performed in the previous subsection but report the behaviour of these two particular clusters. In figure \ref{fig:clusterbyclusteranalysis} we report the result of our analysis for the unweighted scheme. Figure \ref{fig:medianproxy_cluster12_22} reports the value of the ratios along with the errors (black vertical bars). Differently for the whole dataset, we see that from figure \ref{fig:medianproxy_cluster12_22}, the cluster mode removes the vast majority of the memory for cluster 12, without any contribution coming from the market mode or from the interactions. Instead for cluster 22, we see from figure \ref{fig:medianproxy_cluster12_22} that the market is the major contributor to the memory, whereas the cluster mode is reducing some the remaining memory to some extent and the interactions are again not giving much contribution. Figure \ref{fig:ContributionMemory_cluster12_22} reports the same kind of result but relatively to the overall memory. These results suggest that a local analysis reveals a richer behaviour in how the terms in our log volatility factor model affect the memory effect, showing that there is also a link between the correlation structure of the log volatilities and the memory effect. Given these results, we argue that a good criteria for selecting statistically meaningful factors, among all cluster modes, to be included in the definition of our factor model, is to choose those which achieve a significant reduction (in the sense of Section \ref{criterion}) to the memory of the stocks within their cluster. Table \ref{table:ClusterSectorSig} summarizes the results of this procedure, reporting in the first column the cluster number $k$ (as given by the DBHT algorithm). The second column contains the number of stocks in each cluster and in the fourth column we show if the cluster mode reduces the memory of the stocks within that cluster significantly. As we can see we find that out of 29 clusters, 7 do not have a significant meaning to the memory, thus, according, to our criteria are discarded. The fifth, sixth, seventh and eighth column of the table \ref{table:ClusterSectorSig} are the fractional contributions that the market, cluster, interactions and residuals make to the overall memory in the cluster. Comparing the last four columns in table \ref{table:ClusterSectorSig} we see that there is significant heterogeneity in the amount of contributions the market and cluster make to the cluster's overall memory, which highlights the importance of the inclusion of cluster factors in our factor model.
\begin{table}[!htbp]
  \centering
	\begin{adjustbox}{width=0.85\textwidth}
    \begin{tabular}{|c|c|c|c|r|r|r|r|}
    \hline
    k     & no. stocks & dom. supersector  & cluster sig & \multicolumn{1}{c|}{market} & \multicolumn{1}{c|}{cluster} & \multicolumn{1}{c|}{interac} & \multicolumn{1}{c|}{resid} \\
    \hline
    1     & 68    & OG (0) & T     & 0.000 & 0.758 & 0.055 & 0.187 \\
    \hline
    2     & 26    & OG (0) & T     & 0.000 & 0.653 & 0.097 & 0.250 \\
    \hline
    3     & 12    & FS (0) & T     & 0.387 & 0.463 & 0.041 & 0.110 \\
    \hline
    4     & 39    & U (0) & T     & 0.855 & 0.032 & 0.024 & 0.090 \\
    \hline
    5     & 13    & BR (0) & T     & 0.727 & 0.199 & 0.016 & 0.058 \\
    \hline
    6     & 11    & IGS (0.089957) & T     & 0.719 & 0.073 & 0.026 & 0.182 \\
    \hline
    7     & 23    & FS (0) & T     & 0.721 & 0.127 & 0.053 & 0.100 \\
    \hline
    8     & 17    & FB (0) & F     & 0.818 & 0.000 & 0.021 & 0.161 \\
    \hline
    9     & 9     & HC (0) & T     & 0.923 & 0.029 & 0.001 & 0.047 \\
    \hline
    10    & 24    & IGS (0.355912) & T     & 0.471 & 0.403 & 0.028 & 0.098 \\
    \hline
    11    & 11    & HC (0) & F     & 0.890 & 0.000 & 0.018 & 0.093 \\
    \hline
    12    & 32    & RE (0) & T     & 0.000 & 0.977 & 0.005 & 0.018 \\
    \hline
    13    & 30    & FS (0) & T     & 0.662 & 0.226 & 0.019 & 0.093 \\
    \hline
    14    & 144   & RE (0) & T     & 0.574 & 0.272 & 0.049 & 0.105 \\
    \hline
    15    & 77    & HC (0) & T     & 0.769 & 0.093 & 0.012 & 0.127 \\
    \hline
    16    & 5     & TL (0) & T     & 0.968 & 0.012 & 0.003 & 0.016 \\
    \hline
    17    & 66    & B (0) & T     & 0.733 & 0.149 & 0.040 & 0.078 \\
    \hline
    18    & 111   & B (0) & T     & 0.833 & 0.088 & 0.024 & 0.055 \\
    \hline
    19    & 15    & PHG (0) & T     & 0.781 & 0.134 & 0.031 & 0.054 \\
    \hline
    20    & 8     & TL (0) & F     & 0.965 & 0.000 & 0.002 & 0.033 \\
    \hline
    21    & 172   & T (0) & T     & 0.684 & 0.221 & 0.013 & 0.082 \\
    \hline
    22    & 118   & T (0) & T     & 0.836 & 0.071 & 0.020 & 0.073 \\
    \hline
    23    & 14    & I (0) & F     & 0.951 & 0.000 & 0.007 & 0.042 \\
    \hline
    24    & 12    & IGS (0.003514) & T     & 0.911 & 0.050 & 0.005 & 0.034 \\
    \hline
    25    & 17    & C (0) & T     & 0.956 & 0.005 & 0.003 & 0.035 \\
    \hline
    26    & 31    & R (0) & T     & 0.900 & 0.036 & 0.008 & 0.057 \\
    \hline
    27    & 43    & IGS (0) & F     & 0.945 & 0.000 & 0.005 & 0.049 \\
    \hline
    28    & 37    & R (0) & F     & 0.940 & 0.000 & 0.003 & 0.057 \\
    \hline
    29    & 15    & IGS (0) & F     & 0.954 & 0.000 & 0.003 & 0.044 \\
    \hline
    \end{tabular}%
		\end{adjustbox}
		\caption{Table showing the cluster no. k in the first column and the number of stocks in the second column. In the third column, we have the dominant ICB supersector (abbreviated to the first letters in each supersector, which are listed in figure \ref{fig:ClusterSummary}). In brackets in the third column we have the p value of the hypothesis test which tests whether the most dominant supersector can be meaningfully identified from the cluster \cite{Musmeci2015b}, which are given to 6 decimal places. The fourth column details whether the cluster mode significantly reduces the memory for that cluster. The fifth, sixth, seventh and eighth columns are the fraction of contribution (to 3 decimal places) that the market, cluster, interactions and residual make respectively to the total memory.}
  \label{table:ClusterSectorSig}%
\end{table}%
\section{Economical interpretation of the clusters} \label{EconomicalInterpretation}
\begin{figure}
\centering
\includegraphics[width=0.75\textwidth]{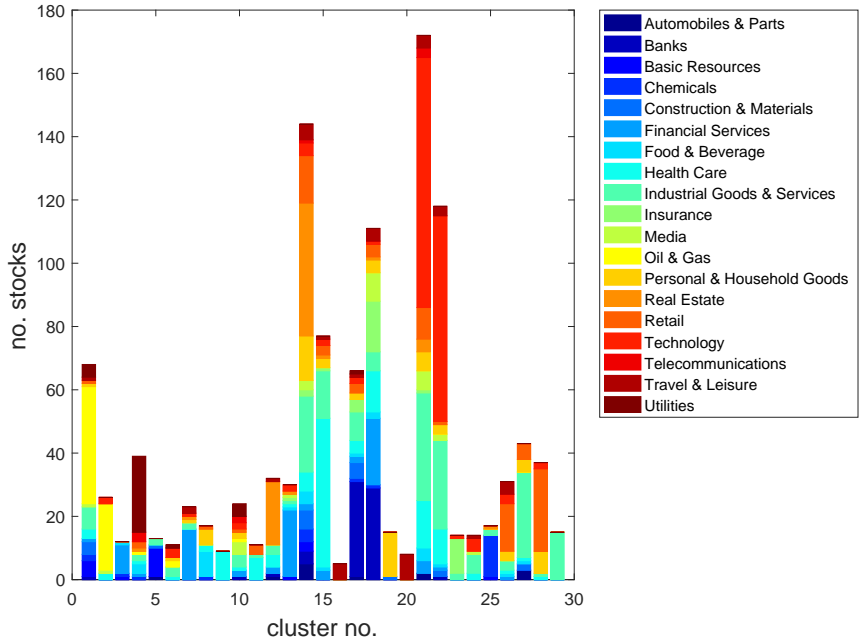}
\caption{Composition of DBHT clusters in terms of ICB supersectors. The x axis labels the clusters of DBHT and the y axis is the number of stocks in each cluster. The colours represent particular ICB supersector given in the key.}
\label{fig:ClusterSummary}
\end{figure}
Up till now, we have focused on determining the clusters via statistical tools. In this section we show that the clusters also have an economical interpretation.  In figure \ref{fig:ClusterSummary}, we show the cluster composition of each cluster identified through DBHT using the Industrial Classification Benchmark (ICB)supersector classification of common industries, with each colour representing a different supersector. In particular from figure \ref{fig:ClusterSummary}, we observe that clusters are dominated by a particular supersector. For example, we see from figure \ref{fig:ClusterSummary} that clusters 12 and 22 show the presence of dominant supersectors: the real estate sector for cluster 12 and technology sector for cluster 22. In order to check that these identifications of dominant sectors are meaningful, we used the same hypothesis test as in \cite{Tumminello2011,Musmeci2015}, which tests the null hypothesis that the cluster has merely randomly been assigned supersector classifications using the hypergeometric distribution versus the alternative hypothesis that the supersector is indeed dominating the cluster. Starting from a significance level of 5\%, we additionally used a conservative Bonferroni correction for multiple hypothesis testing \cite{Feller2008} of $0.5 N_{cl}N_{ICB}$ to reduce the level of significance, where $N_{cl}$ is the number of clusters identified through DBHT and $N_{ICB}$ is the number of ICB supersectors. This reduces the level of significance to $9.0\times 10^{-5}$, reporting the p values to six decimal places. Table \ref{table:ClusterSectorSig} details the results of applying this process to all clusters, with the dominant supersector denoted in the third column. We see from Table \ref{table:ClusterSectorSig} that in $26$ clusters, the cluster can indeed be matched to their dominating supersector, and of the clusters that significantly contribute to their own memory (see section \ref{ClusterByCluster}), $19$ correspond to their dominating supersector. This opens the possibility of choosing cluster modes for a further refinement of the factor model between log volatilties by choosing the cluster modes which reduce the memory statistically significantly after the market mode is removed, but also having an economic interpretation of being dominated by particular supersectors.

Moreover, after comparing clusters which are dominated by the same ICB supersector in table \ref{table:ClusterSectorSig}, we see that the groups of clusters $k=1,2$ and $k=17,18$, which are dominated by the Oil and Gas and Banks supersectors respectively, have similar contributions for the market, clusters and interactions. However, there are instances where clusters dominated by the same supersector do not have similar contributions. For example, clusters $k=12,14$ are both dominated by the Real Estate supersector, but for $k=12$ the market does not statistically contribute to the memory, whilst for $k=14$ it does. This could be indication of markets moving away from clearly defined industrial supersectors, which was also noted in \cite{Musmeci2015b}, and emphasises why we have used the clustering algorithm DBHT, rather than taking the industrial classifications directly.
\section{Comparison with PCA and Exploratory Factor Analysis} \label{ComparisonPCA}
\begin{figure}
\centering
\includegraphics[width=0.9\textwidth]{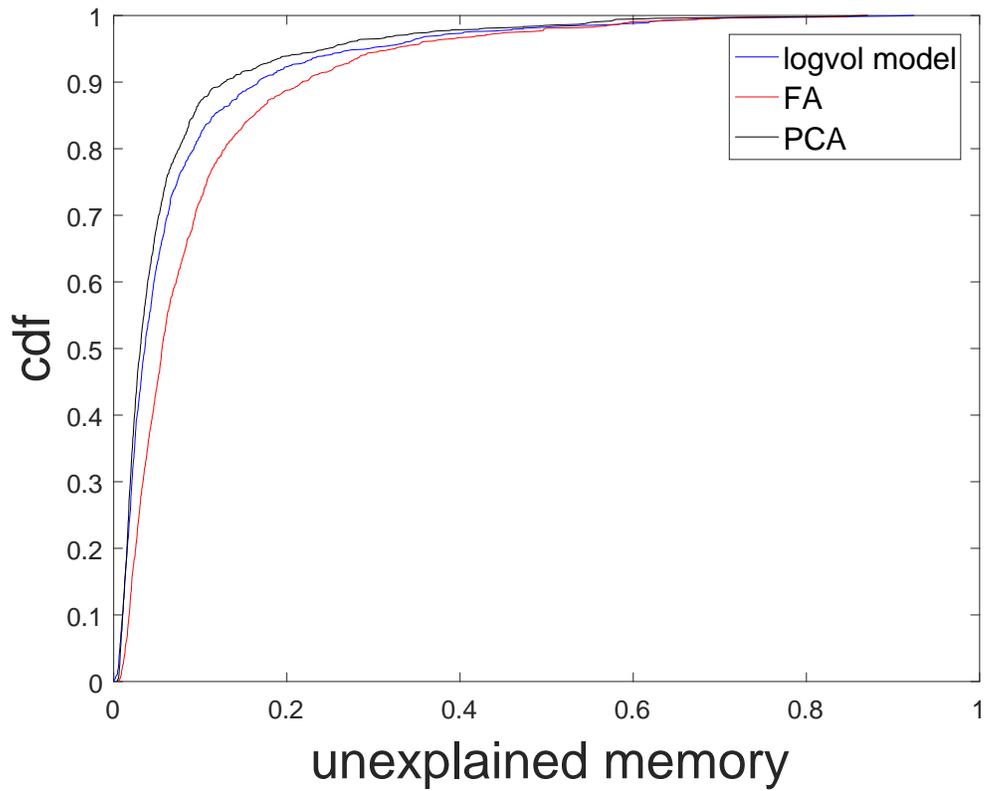}
\caption{Empirical cumulative distribution function of the unexplained residual memory for the factor model in blue line, the PCA in black, where we only take the first 23 principal components, and the exploratory factor analysis, where we use 23 factors and a varimax rotation.}
\label{fig:CDF_residualcomparison}
\end{figure}

In this section we compare the memory reduction performance of our model with a well established PCA inspired factor model \cite{Jolliffe1982} and exploratory factor analysis driven factor model. Firstly, we explain the importance of the PCA factor model. The PCA analysis gives a set of orthogonal eigenvectors that define mutually linearly uncorrelated portfolios that can be used to help define factor models by assigning each principal component a separate factor. However, as we have pointed out it is difficult to decide how many principal components we should keep. In our analysis, the number of principal components we keep in the PCA factor model shall be fixed to be the same as the number of factors in our factor model i.e. 20. PCA aims to explain the diagonal terms in the orthogonal basis of the correlation matrix $\mathbf{E}$, which is the correlation matrix between the $\ln |r_{i}(t)|$. Exploratory factor analysis (FA) on the other hand is more general, and aims to explain the off diagonal terms of $\mathbf{E}$, using the general linear model in \eqref{GLM}. Again, there are problems selecting exactly how many factors should be included \cite{Preacher2013}, but we fix the number of factors in the FA model to be equal to the number of factors in our log volatility factor model i.e. 20. After extracting the factors, we apply a varimax rotation of the factors \cite{Child2006}, which is commonly applied in factor analysis to improve understandability. In figure \ref{fig:CDF_residualcomparison} we plot the cumulative distribution function of how much residual memory is left after removal of the factors for the log volatility factor model, FA model and PCA factor model as a percentage of the total memory before removal.

We see from figure \ref{fig:CDF_residualcomparison} that 90\% of all stocks only have a maximum of 16.7\% residual memory left for the factor model of log volatility, whereas 90\% of all stocks have a maximum of 12.7\% of residual memory left, which means that the PCA factor model and the log volatility factor model both explain the memory to the same efficiency. For the exploratory factor model, we see that 90\% of all stocks have 21.8\% of their memory left, which is worse than the log volatility factor model and the PCA factor model, but still has a comparable performance. We can therefore conclude that the log volatility factor model explains the same amount of memory as the other two models, even after fixing the amount of factors to be the same in the PCA and exploratory factor model.     

\section{Dynamic Stability of Clusters and their Memory Properties} \label{DynamicalStability}
\begin{figure}
\centering
\includegraphics[width=0.9\textwidth]{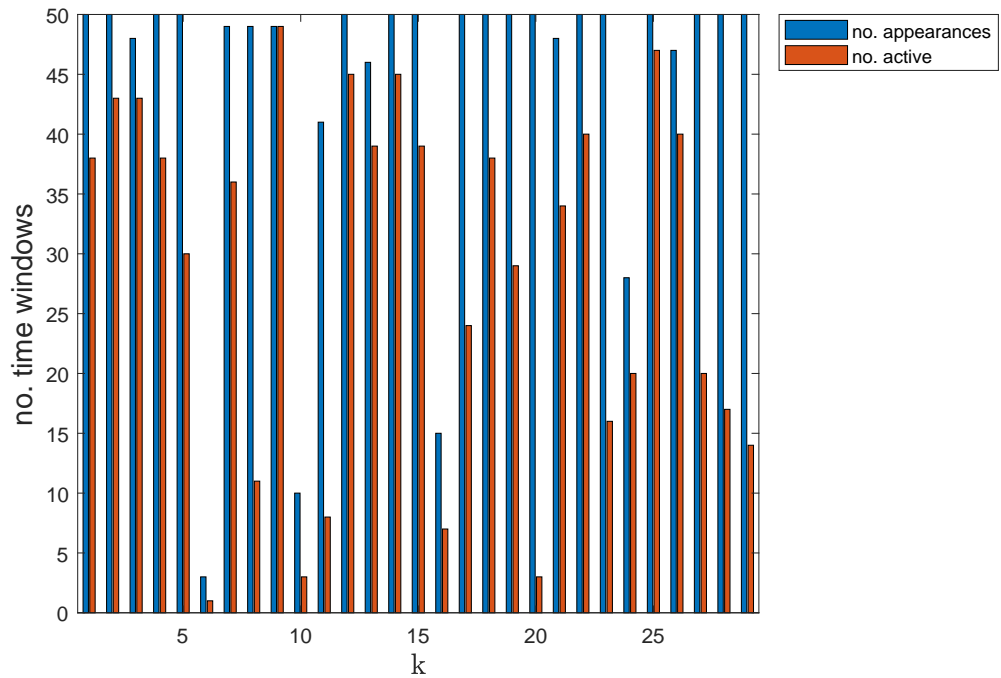}
\caption{The blue bars are the number of time windows where a cluster $k$ (whose identity are detailed in \ref{table:ClusterSectorSig}), can be statistically identified with a cluster in $Y^{m}$, which is the clustering computed over $50$ rolling windows of length $1600$. The red bars are the number of time windows where a cluster in $X$ can statistically reduce its own memory on the rolling time window $m$.}
\label{fig:DynamicalStability}
\end{figure}

So far the results that have been presented are based on static correlation matrices, which are computed across the whole time period considered in our dataset. A natural question then arises about whether the results presented in sections \ref{MemoryFiltration} and \ref{EconomicalInterpretation} are dynamically stable. First we divide each stock's time series into $50$ rolling windows of length $1600$, which gives a shift of $56$ days for each window \cite{Musmeci2015b}. For every window, we then perform the same analysis as is done in section \ref{NewLogVolFactorModel}. That is, for each rolling window $m=1,2,...,50$ we remove the market mode computed on that time window, and then compute the corresponding correlation matrix $\mathbf{G}^{m}$ and its clustering $Y^{m}$ using the DBHT algorithm. To tackle whether the clusters themselves are dynamically stable, we use a similar procedure to the one presented in section \ref{EconomicalInterpretation} and is also carried out in \cite{Musmeci2015b}. Specifically, for each time window $m$, we use the hypergeometric test to see if each of the clusters in the static clustering $X$ are statistically similar to a cluster in $Y^{m}$, recording the number of time windows where there is a possible match. This is recorded in the blue bars in figure \ref{fig:DynamicalStability}. We also calculate whether each of the clusters in $Y^{m}$ can still statistically reduce their memory in every time window $m$, and measure the total number of time windows where this happens, which is plotted in the red bars in figure \ref{fig:DynamicalStability}. These two numbers give a measure of persistence of both the appearance and statistical memory reduction properties for each cluster.

As we can see from figure \ref{fig:DynamicalStability} most clusters are quite stable, appearing in most time windows. The exceptions to this are clusters $6$, $10$, $24$, which interestingly can all be identified with the Industrial Goods and Services supersector (from table \ref{table:ClusterSectorSig}), and cluster $16$, which is quite a small cluster with only $5$ stocks and thus more likely to be unstable in time due its small size. From figure \ref{fig:DynamicalStability}, we can also conclude that the memory filtration properties of the clusters identified in section \ref{ClusterByCluster} are stable in time. This is because figure \ref{fig:DynamicalStability} indicates high persistence of the static clusters from table \ref{table:ClusterSectorSig} in the memory sense (red bars) that statistically contribute to their memory (for example clusters 2 and 12). On the other hand, we see low persistence in the memory sense for clusters that do not contribute to their own memory in the static case (for example clusters 27,28,29). 
\section{Conclusion} \label{Conclusion}

We proposed a new factor model for the log-volatility discussing how each term of the model affects the stylized fact of the volatility clustering. This reduces the information present in the linear correlation between the log volatilities to a global factor, which is the so-called market mode, and second and third local factors, which are the cluster mode and the interactions. Using a new non parametric, integrated proxy for the volatility clustering, we found that there is indeed a link between the volatility and volatility clustering. First, the dataset was examined globally, which revealed the market to account for the majority of the volatility clustering effect present in our dataset. However, a local cluster by cluster analysis instead reveals significant variability: in some clusters, the cluster mode itself may be contributing to the volatility clustering. This enabled us to select only statistically relevant cluster factors, reducing the information in the correlation between the log volatility and the number of factors further. From these reduced set of factors, we can select factors that have an economic interpretation through the identification of their dominant ICB supersector, which decreased the number of relevant factors some more. This is significantly advantageous over other potential factor models that could be used for log volatility such as PCA and exploratory factor analysis since we do not subjectively select the number of factors, and also because the factors have a clearer economic interpretation through the identification of their dominant ICB supersector. A comparison of the log volatility factor model with PCA and an exploratory factor model reveals that they explain the same amount of memory in the dataset. Both the clusters and their reported memory filtration were also found to be dynamically stable. 

This work is particularly relevant for the field of volatility modelling, since most multivariate models such as multivariate extensions of GARCH, stochastic covariance and realised covariance models suffer from the curse of dimensionality and increase in the number of parameters. The log volatility factor model presented here could be used to help reduce the amount of parameters needed for these models through the identification of a reduced set of factors given by the procedure in this paper.

\appendix
\section{Appendix}
\subsection{Data cleaning process} \label{DataCleaning}
Our dataset cannot be used as it is since the price time-series are not aligned, which is due to the fact the some stocks have not been traded on certain days. In order to overcome this issue, we apply a data cleaning procedure which allows us to keep as many stock as possible. For example, we do not want to remove a stock just because it was not traded on few days in the given time-span. The main idea is to fill the gaps dragging the last available price and assuming that a gap in the price time-series corresponds to a zero log-return. At the same time we do not want to drag too many prices because a time-series filled with zeros would not be statistically significant. In light of this we remove from our dataset the time-series which are too short in a certain sense. The detailed procedure goes as follows:
\begin{enumerate}
\item Remove from the dataset the price time-series with length less than $p$ times the longest one;
\item Find the common earliest day among the remaining time-series;
\item Create a reference time-series of dates when at least one of the stocks has been traded starting from the earliest common date found in the previous step;
\item Compare the reference time-series of dates with the time-series of dates of each stock and fill the gaps dragging the last available price.
\end{enumerate}
In this paper we chose $p=0.90$ thus keeping as much as possible unmodified time-series. However, the results do not change if we pick a higher value of $p$.

\subsection{Weighting schemes} \label{WeightingSchemes}
Here we shall define the two types of weighting schemes used in this paper for the $\xi_{i}$ and $\xi_{ik}$ defined in \eqref{MarketModeWeighted} and \eqref{ClusterModeWeighted} respectively. The first weighting scheme is based on the eigenspectrum of $\mathbf{E}$ and $\mathbf{G}$. It is useful now to explain the financial interpretation of the eigenvectors $\mathbf{v}$ with entries $v_{i}$ and eigenvalue $\lambda$ for $\mathbf{E}$. $v_{i}$ can be seen as weights for a portfolio defined by $\mathbf{v}$. Measuring the risk from the volatility of the portfolio via its variance, we see it is given by:
\begin{equation}
\frac{1}{T}\sum_{t}\left(\sum_{i}v_{i}\ln |r_{i}(t)|\right)^{2}=\sum_{ij}v_{i}v_{j}E_{ij}=\lambda
\end{equation}
Hence $\lambda$ represents the risk from the volatility of the portfolio given by $\mathbf{v}$. We set $\xi_{i}=v_{i}$, where now $v_{i}$ is the $ith$ entry of the eigenvector corresponding to the largest eigenvalue of the empirical correlation matrix $\mathbf{E}$. This is called the market eigenvalue as it represents all stocks moving together \cite{Plerou2002}, and is also portfolio of stocks that gives the risk of the market volatility mode through its corresponding eigenvalue. We could have also used a real index to determine the weights e.g. the Dow Jones, but \cite{Borghesi2007} showed that this does not effectively remove the influence of modes from returns compared to a pseudo-index.

The weights $\xi_{ik}$ are established in a similar way to the market mode case, which we shall do by considering only the part of $\mathbf{G}$ which corresponds to members of the cluster. Defining a submatrix of $\mathbf{G}$
\begin{equation}
\mathbf{G}^{(k)}=\left\{\mathbf{G}\right\}_{(i,j)\in cluster  k}
\end{equation} 
Where $\left\{...\right\}_{(i,j)\in cluster \ k}$ refers to only keeping the elements the matrix in which $i$ and $j$ are stocks in cluster $k$. Thus $\mathbf{G}^{(k)}$ is the square sub matrix of $\mathbf{G}$ corresponding to cluster $k$. This submatrix is the correlation matrix of a market which consists only of stocks which are part of cluster $k$. Hence, in exactly the same way as the market eigenvalue, the largest eigenvalue of $\mathbf{G}^{(k)}$ represents stocks of the cluster moving together, the value of the eigenvalue being the risk of the cluster market portfolio, and the related eigenvector giving the weights of such a portfolio. Therefore, the definition of the weights $\xi_{ik}$ for cluster $k$ are determined by setting $\omega_{ik}=v_{i}^{(k)}$, which is the $ith$ entry of the eigenvector corresponding to the largest eigenvalue of $\mathbf{G}^{(k)}$. This is the weighting scheme used and is compared to the case of equal weights where $\xi_{i}=\frac{1}{N}$ and $\xi_{ik}=\frac{1}{m_{k}}$ in figures \ref{fig:medianvolproxy}, \ref{fig:contributionmarket} and \ref{fig:cumintproxy} thereafter the equal weights scheme is used.

\subsection{Elastic Net Regression} \label{ElasticNet}
Elastic net regression is used to find the values of $\beta_{ik}$ and $\beta_{ik'}$ using Eq. \eqref{Residue}. Further details of the use of this method is provided in this appendix. Elastic net regression \cite{Zou2005} is a hybrid version of ridge regularisation and lasso regression, thus providing a way of dealing with correlated explanatory variables (in our case $I_{k}(t)$ and $I_{k'}(t)$) and also performing feature selection, which takes into account non-interacting clusters $I_{k'}(t)$ that ridge regularisation would ignore. Elastic net regression solves the constrained minimisation problem
\begin{equation}
\min_{\boldsymbol{\beta}_{i}} \frac{1}{T}\sum_{t=1}^{T}\left(c_{i}(t)-\mathbf{I}(t)^{\dagger}\boldsymbol{\beta}_{i}\right)^{2}+\lambda P_{a}(\boldsymbol{\beta}_{i})
\end{equation}
, where $\boldsymbol{\beta}_{i}$ is the vector of loadings given by $(\beta_{i1}, \beta_{i2}, \dots,\beta_{iK})^{\dagger}$, $\mathbf{I}(t)$ is the matrix consisting of columns $(I_{1}(t),I_{2}(t), \dots, I_{N_{cl}}(t))$ and $\lambda$ and $a$ are hyperparameters. $P_{a}(\boldsymbol{\beta}_{i})$ is defined as
\begin{equation}
P_{a}(\boldsymbol{\beta}_{i})=\sum_{j=1}^{M}\left((1-a)\frac{\beta_{ij}^{2}}{2}+a |\beta_{ij}|\right) \label{ElasticNetPenalty}
\end{equation}  
. The first term in the sum of Eq. \eqref{ElasticNetPenalty} is the $L_{2}$ penalty for the ridge regularisation and the second term in the sum is the $L_{1}$ penalty for the lasso regression. Hence if $a=0$ then elastic net reduces to ridge regression and if $a=1$ then elastic net becomes lasso, with a value between the two controlling the extent which one is preferred to the other. The determination of the $a$ hyperparameter, controlling the extent of lasso vs ridge, and $\lambda$, for the ridge, is done using 10 cross validated fits \cite{Zou2005}, picking the pair of $(a,\lambda)$ that give the minimum prediction error. We show the values of $\beta_{ik}$ and test the significance of the predictor $I_{k}(t)$ at the 5\% level in Table \ref{tab:PValElasticNet} , where the p value is shown in brackets, using the significance test outlined in \cite{Lockhart2014}.
\begin{table}[htbp]
  \centering
    \begin{tabular}{|l|l|l|}
    \hline
          & \multicolumn{2}{c|}{$\beta_{ik}$} \\
    \hline
    KO    & 0.9431(0) & 0.8997(0) \\
    \hline
    RIG   & 0.9041 (0) & 1.1265(0) \\
    \hline
    \end{tabular}%
	\caption{This table shows the responsiveness to the cluster mode $I_{k}(t)$, $\beta_{ik}$ calibrated as detailed in section \ref{ClusterModeCal}. P values shown in brackets test the significance of the predictor given by the cluster mode $I_{k}(t)$. The first column is for the weighted scheme and second is for equal weights, which are detailed in \ref{WeightingSchemes}.}
	\label{tab:PValElasticNet}%
\end{table}%

\subsection{Visualisation of Residuals and Factors}
\begin{figure}
\centering
\includegraphics[width=0.75\textwidth]{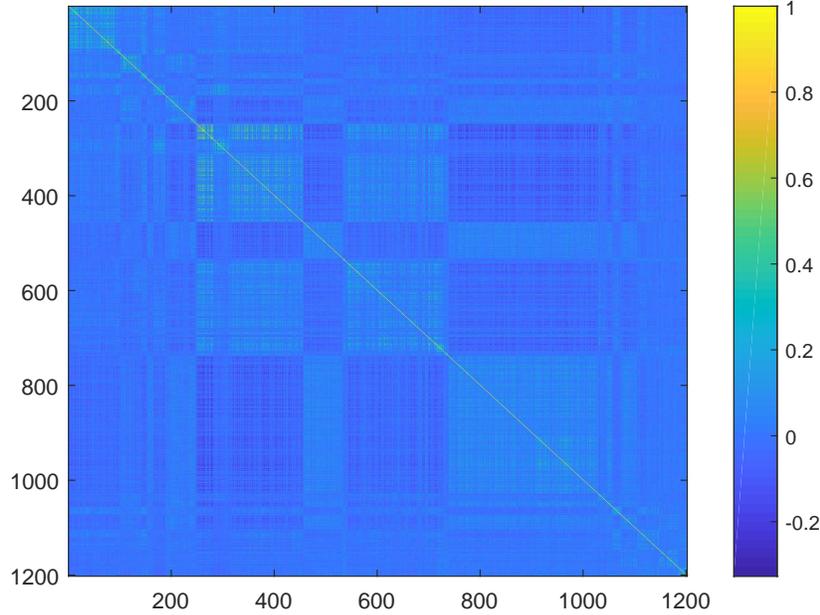}
\caption{Heat map of the correlation matrix for $\mathbf{G}$ with the stocks reordered to correspond to their cluster no. from table \ref{table:ClusterSectorSig}. The colour legend for the heat map is given to the right of the figure.}
\label{fig:CorrMatrix_HeatMap}
\end{figure}

We can represent the correlation matrix $\mathbf{G}$ defined in eq. \eqref{CorrResidue} as a heat map, which is shown in figure \ref{fig:CorrMatrix_HeatMap} with the stocks reordered according their cluster no. $k$ given by table \ref{table:ClusterSectorSig}. From figure \ref{fig:CorrMatrix_HeatMap}, we see the clusters of correlation matrix, which are given by the square blocks along the diagonal that are more populated by higher correlation values. We also see the interactions between the clusters which are represented by the rectangular blocks of higher correlation values away from the main diagonal. 

\begin{figure}
\centering
\begin{subfigure}{0.475\textwidth}
\centering
\includegraphics[width=\textwidth]{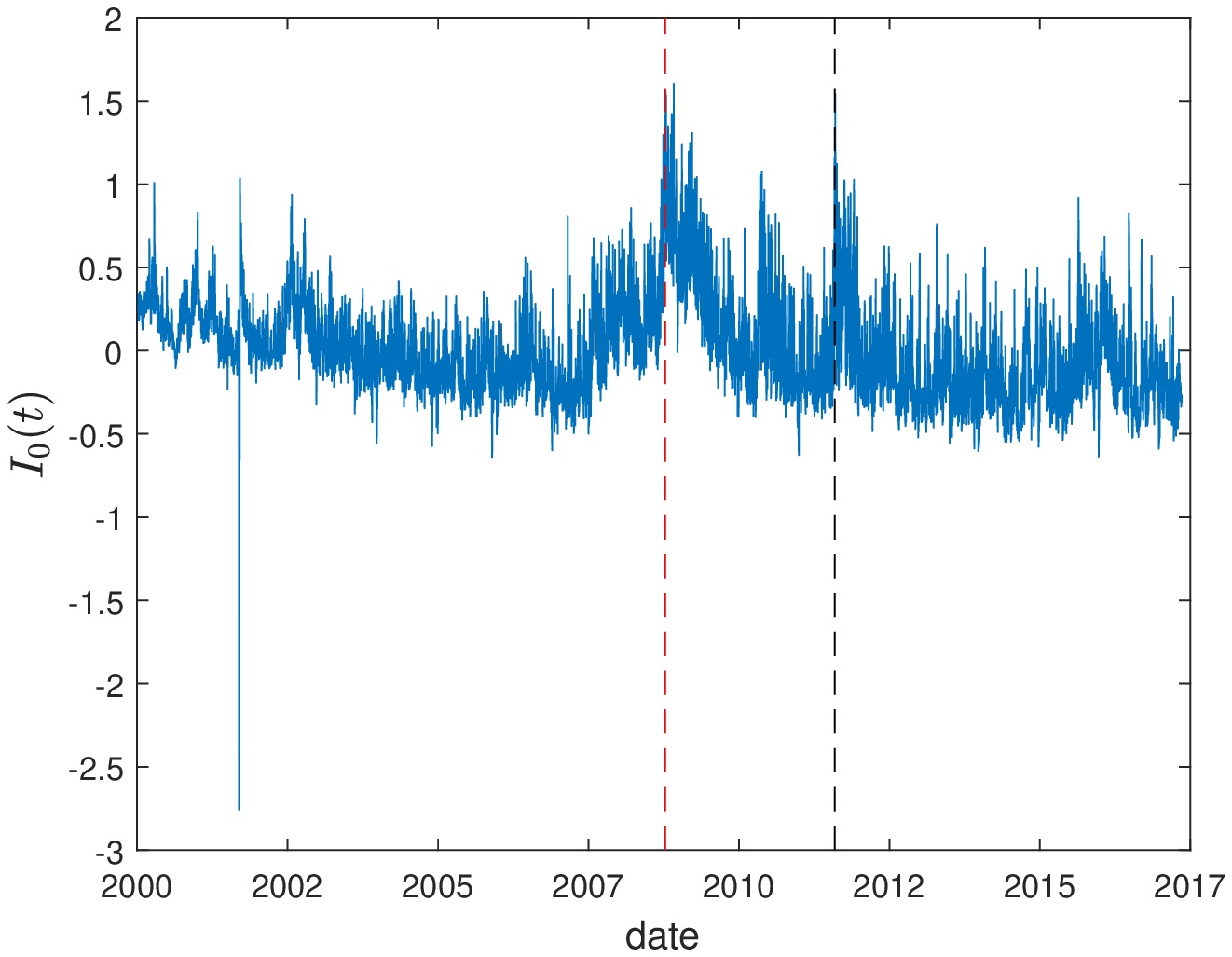}
\caption{$I_{0}(t)$}
\label{fig:MarketMode}
\end{subfigure} \\
\begin{subfigure}{0.475\textwidth}
\centering
\includegraphics[width=\textwidth]{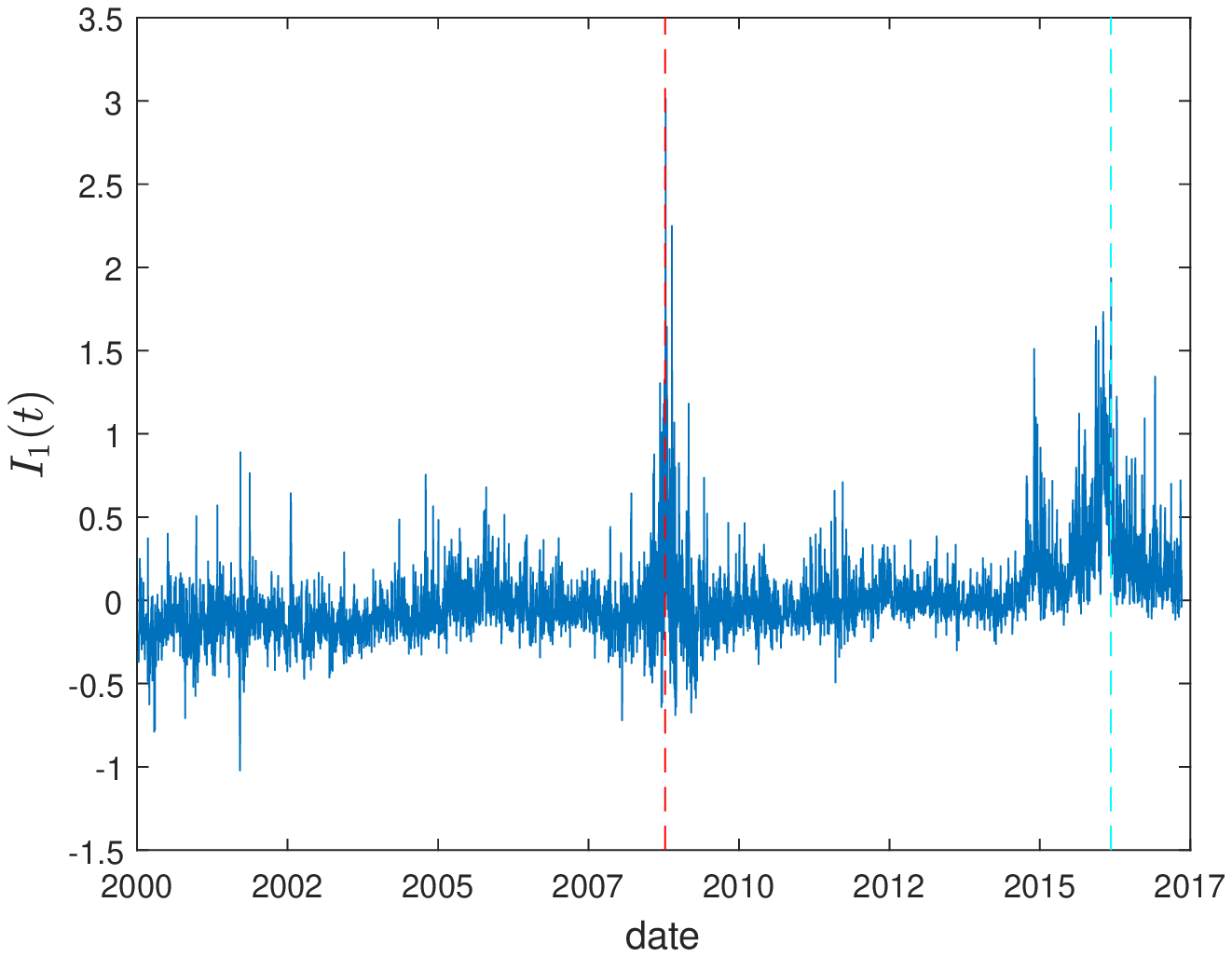}
\caption{$I_{1}(t)$}
\label{fig:ClusterMode_1}
\end{subfigure}
\begin{subfigure}{0.475\textwidth}
\centering
\includegraphics[width=\textwidth]{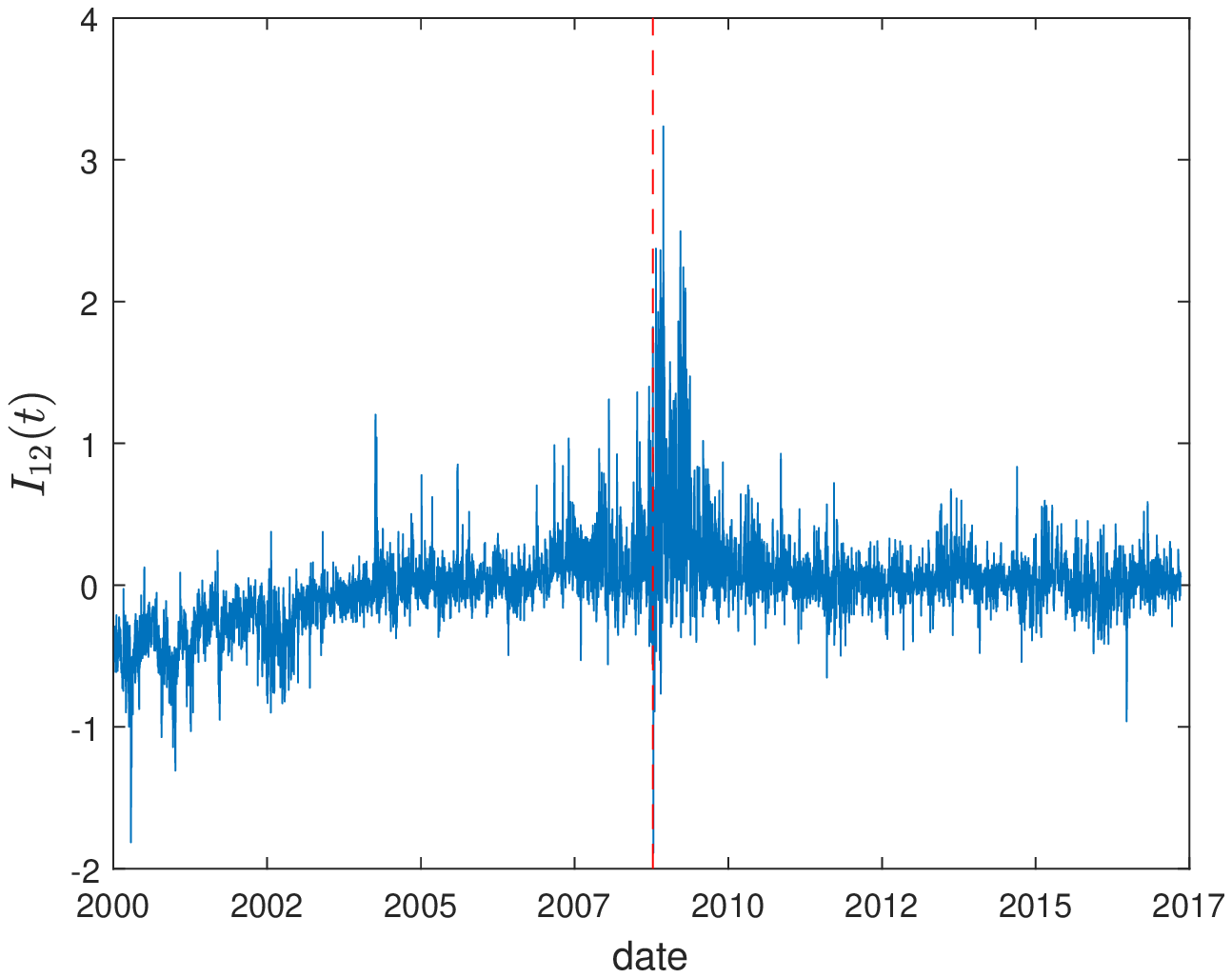}
\caption{$I_{12}(t)$}
\label{fig:ClusterMode_12}
\end{subfigure}
\caption{Time series of the market mode $I_{0}(t)$ in \subref{fig:MarketMode}, and the cluster modes $I_{k}(t)$ for $k=1,12$ (see table \ref{table:ClusterSectorSig}) respectively in \subref{fig:ClusterMode_1} and \subref{fig:ClusterMode_12}, where the subscripts of the cluster modes refers to the clusters given in table \ref{table:ClusterSectorSig}. The red dashed lines in these plots refers to the outbreak of the Great Financial Crisis of 2008. The black dashed line in figure \ref{fig:MarketMode} marks a portion of the Eurozone debt crisis. The light blue dashed line in figure \ref{fig:ClusterMode_1} marks a low global demand in oil and gas supplies. }
\label{fig:timeseries_modes}
\end{figure}

In order to provide a visualisation of the factors, we plot the time series of the market mode $I_{0}(t)$ and the two particular cluster modes $I_{k}(t)$ for $k=1,12$, where the subscript of the cluster modes indicates the particular clusters we are using from table \ref{table:ClusterSectorSig}, in figure \ref{fig:timeseries_modes}. We see from figure \ref{fig:timeseries_modes} that the time series encodes important information regarding market conditions. In the plot for $I_{0}(t)$ in figure \ref{fig:MarketMode}, the two periods of high volatility indicated by the red and black dashed lines represent the Great Financial Crisis of 2008 and the Eurozone Debt Crisis (note that the extreme low volatility seen before 2002 was caused by the American stock exchanges being shut down due to the September 11th terrorist attack). The time series of $I_{1}(t)$ in figure \ref{fig:ClusterMode_1} again shows a high volatility period during the financial crisis, but we also see another high volatility phase denoted by the light blue dashed line. This represents the volatility in the oil and gas markets caused by low demand, and makes sense since table \ref{table:ClusterSectorSig} shows that cluster 1 represents the Oil and Gas ICB supersector.

\subsection{Smoothness of $\eta$}
\begin{figure}
\centering
\begin{subfigure}{0.475\textwidth}
	\centering
	\includegraphics[width=\textwidth]{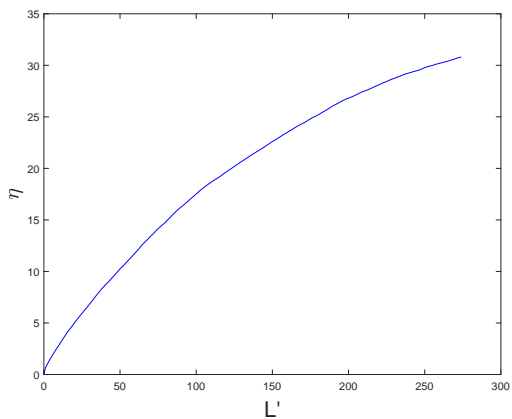}
	\caption{Coca Cola Enterprises Inc. }
	\label{fig:IntProxy_KO}
\end{subfigure}
\hfill
\begin{subfigure}{0.475\textwidth}
	\centering
	\includegraphics[width=\textwidth]{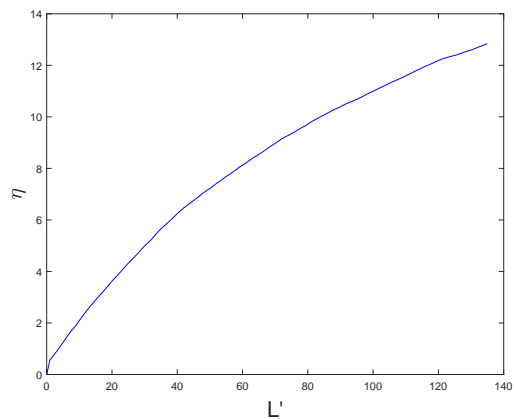}
	\caption{Transoceanic}
	\label{fig:IntProxy_RIG}
\end{subfigure}
\caption{Integrated proxy $\eta$ as a function of the lag $L'$ where $\eta$ is integrated over [1,L'] until $L'=L_{cut}$. Fig. \ref{fig:IntProxy_KO} is for Coca Cola Co. and fig. \ref{fig:IntProxy_RIG} for Transocean}
\label{fig:IntProxy_Examples}
\end{figure}

We plot $\eta$ as a function of the upper limit in the integrand of eq. \eqref{IntProxy}, where the upper limit $L'$ is allowed to be in the interval $[1,L_{cut}]$. As we can see from both plots in figure \ref{fig:IntProxy_Examples}, the line is much smoother showing that the $\eta$ proxy is much more robust with respect to the noisy signal of the empirical ACF. This offers an advantage of using $\eta$ rather than $\beta^{\text{vol}}$ which is more sensitive the the noise in the ACF and gives poor fits to the ACF in log-log scale as can be seen from the examples in figure \ref{fig:ACFExamples}. 

\bibliographystyle{unsrt}
\bibliography{BibTeX1}

\begin{thebibliography}{10}

\bibitem{Bouchaud2009b}
Jean-Philippe Bouchaud and Marc Potters.
\newblock {\em Theory of financial risk and derivative pricing: from
  statistical physics to risk management}.
\newblock Cambridge university press, 2009.

\bibitem{Hull1987}
John Hull and Alan White.
\newblock The pricing of options on assets with stochastic volatilities.
\newblock {\em The journal of finance}, 42(2):281--300, 1987.

\bibitem{Hull2006}
John~C Hull.
\newblock {\em Options, futures, and other derivatives}.
\newblock Pearson Education India, 2006.

\bibitem{Bun2017}
Jo{\"e}l Bun, Jean-Philippe Bouchaud, and Marc Potters.
\newblock Cleaning large correlation matrices: tools from random matrix theory.
\newblock {\em Physics Reports}, 666:1--109, 2017.

\bibitem{Bauwens2006}
Luc Bauwens, S{\'e}bastien Laurent, and Jeroen~VK Rombouts.
\newblock Multivariate garch models: a survey.
\newblock {\em Journal of applied econometrics}, 21(1):79--109, 2006.

\bibitem{Clark1973}
Peter~K Clark.
\newblock A subordinated stochastic process model with finite variance for
  speculative prices.
\newblock {\em Econometrica: journal of the Econometric Society}, pages
  135--155, 1973.

\bibitem{Andersen2003}
Torben~G Andersen, Tim Bollerslev, Francis~X Diebold, and Paul Labys.
\newblock Modeling and forecasting realized volatility.
\newblock {\em Econometrica}, 71(2):579--625, 2003.

\bibitem{Maaten2009}
Laurens Van Der~Maaten, Eric Postma, and Jaap Van~den Herik.
\newblock Dimensionality reduction: a comparative.
\newblock {\em J Mach Learn Res}, 10:66--71, 2009.

\bibitem{Jolliffe1986}
Ian~T Jolliffe.
\newblock Principal component analysis and factor analysis.
\newblock In {\em Principal component analysis}, pages 115--128. Springer,
  1986.

\bibitem{Darbyshire2017}
J~Darbyshire.
\newblock {\em The volatility surface: a practitioner's guide}, volume 357.
\newblock Aitch \& Dee Limited, 2017.

\bibitem{Alexander2002}
Carol Alexander.
\newblock Principal component models for generating large garch covariance
  matrices.
\newblock {\em Economic Notes}, 31(2):337--359, 2002.

\bibitem{Zhang2009}
Kun Zhang and Laiwan Chan.
\newblock Efficient factor garch models and factor-dcc models.
\newblock {\em Quantitative Finance}, 9(1):71--91, 2009.

\bibitem{Plerou2002}
Vasiliki Plerou, Parameswaran Gopikrishnan, Bernd Rosenow, Luis A~Nunes Amaral,
  Thomas Guhr, and H~Eugene Stanley.
\newblock Random matrix approach to cross correlations in financial data.
\newblock {\em Physical Review E}, 65(6):066126, 2002.

\bibitem{Majumdar2012}
Satya~N Majumdar and Pierpaolo Vivo.
\newblock Number of relevant directions in principal component analysis and
  wishart random matrices.
\newblock {\em Physical review letters}, 108(20):200601, 2012.

\bibitem{Jackson1993}
Donald~A Jackson.
\newblock Stopping rules in principal components analysis: a comparison of
  heuristical and statistical approaches.
\newblock {\em Ecology}, 74(8):2204--2214, 1993.

\bibitem{Livan2011}
Giacomo Livan, Simone Alfarano, and Enrico Scalas.
\newblock Fine structure of spectral properties for random correlation
  matrices: An application to financial markets.
\newblock {\em Physical Review E}, 84(1):016113, 2011.

\bibitem{Sharpe1964}
William~F Sharpe.
\newblock Capital asset prices: A theory of market equilibrium under conditions
  of risk.
\newblock {\em The journal of finance}, 19(3):425--442, 1964.

\bibitem{Roll1980}
Richard Roll and Stephen~A Ross.
\newblock An empirical investigation of the arbitrage pricing theory.
\newblock {\em The Journal of Finance}, 35(5):1073--1103, 1980.

\bibitem{Fama1993}
Eugene~F Fama and Kenneth~R French.
\newblock Common risk factors in the returns on stocks and bonds.
\newblock {\em Journal of financial economics}, 33(1):3--56, 1993.

\bibitem{Chicheportiche2015}
R{\'e}my Chicheportiche and J-P Bouchaud.
\newblock A nested factor model for non-linear dependencies in stock returns.
\newblock {\em Quantitative Finance}, 15(11):1789--1804, 2015.

\bibitem{Fama1996}
Eugene~F Fama and Kenneth~R French.
\newblock Multifactor explanations of asset pricing anomalies.
\newblock {\em The journal of finance}, 51(1):55--84, 1996.

\bibitem{Engel2015}
Charles Engel, Nelson~C Mark, and Kenneth~D West.
\newblock Factor model forecasts of exchange rates.
\newblock {\em Econometric Reviews}, 34(1-2):32--55, 2015.

\bibitem{Thompson2004}
Bruce Thompson.
\newblock {\em Exploratory and confirmatory factor analysis: Understanding
  concepts and applications.}
\newblock American Psychological Association, 2004.

\bibitem{Merton1973}
Robert~C Merton.
\newblock An intertemporal capital asset pricing model.
\newblock {\em Econometrica: Journal of the Econometric Society}, pages
  867--887, 1973.

\bibitem{Zabarankin2014}
Michael Zabarankin, Konstantin Pavlikov, and Stan Uryasev.
\newblock Capital asset pricing model (capm) with drawdown measure.
\newblock {\em European Journal of Operational Research}, 234(2):508--517,
  2014.

\bibitem{Barberis2015}
Nicholas Barberis, Robin Greenwood, Lawrence Jin, and Andrei Shleifer.
\newblock X-capm: An extrapolative capital asset pricing model.
\newblock {\em Journal of Financial Economics}, 115(1):1--24, 2015.

\bibitem{Markowitz1952}
Harry Markowitz.
\newblock Portfolio selection.
\newblock {\em The journal of finance}, 7(1):77--91, 1952.

\bibitem{Fama1992}
Eugene~F Fama and Kenneth~R French.
\newblock The cross-section of expected stock returns.
\newblock {\em the Journal of Finance}, 47(2):427--465, 1992.

\bibitem{Connor2012}
Gregory Connor, Matthias Hagmann, and Oliver Linton.
\newblock Efficient semiparametric estimation of the fama--french model and
  extensions.
\newblock {\em Econometrica}, 80(2):713--754, 2012.

\bibitem{Faff2014}
Robert Faff, Philip Gharghori, and Annette Nguyen.
\newblock Non-nested tests of a gdp-augmented fama--french model versus a
  conditional fama--french model in the australian stock market.
\newblock {\em International Review of Economics \& Finance}, 29:627--638,
  2014.

\bibitem{Fama2015}
Eugene~F Fama and Kenneth~R French.
\newblock A five-factor asset pricing model.
\newblock {\em Journal of Financial Economics}, 116:1--22, 2015.

\bibitem{Chen1986}
Nai-Fu Chen, Richard Roll, and Stephen~A Ross.
\newblock Economic forces and the stock market.
\newblock {\em Journal of business}, pages 383--403, 1986.

\bibitem{Reinganum1981}
Marc~R Reinganum.
\newblock The arbitrage pricing theory: some empirical results.
\newblock {\em The Journal of Finance}, 36(2):313--321, 1981.

\bibitem{Faff2004}
Robert Faff.
\newblock A simple test of the fama and french model using daily data:
  Australian evidence.
\newblock {\em Applied Financial Economics}, 14(2):83--92, 2004.

\bibitem{Grauer2010}
Robert~R Grauer and Johannus~A Janmaat.
\newblock Cross-sectional tests of the capm and fama--french three-factor
  model.
\newblock {\em Journal of banking \& Finance}, 34(2):457--470, 2010.

\bibitem{Racicot2016}
Fran{\c{c}}ois-Eric Racicot and William~F Rentz.
\newblock Testing fama--french’s new five-factor asset pricing model:
  evidence from robust instruments.
\newblock {\em Applied Economics Letters}, 23(6):444--448, 2016.

\bibitem{Malevergne2004}
Yannick Malevergne and D~Sornette.
\newblock Collective origin of the coexistence of apparent random matrix theory
  noise and of factors in large sample correlation matrices.
\newblock {\em Physica A: Statistical Mechanics and its Applications},
  331(3):660--668, 2004.

\bibitem{Tumminello2007}
Michele Tumminello, Fabrizio Lillo, and Rosario~N Mantegna.
\newblock Hierarchically nested factor model from multivariate data.
\newblock {\em EPL (Europhysics Letters)}, 78(3):30006, 2007.

\bibitem{Song2012}
Won-Min Song, Tiziana Di~Matteo, and Tomaso Aste.
\newblock Hierarchical information clustering by means of topologically
  embedded graphs.
\newblock {\em PLoS One}, 7(3):e31929, 2012.

\bibitem{Musmeci2015}
Nicolo Musmeci, Tomaso Aste, and Tiziana Di~Matteo.
\newblock Relation between financial market structure and the real economy:
  comparison between clustering methods.
\newblock {\em PloS one}, 10(3):e0116201, 2015.

\bibitem{Taylor1994}
Stephen~J Taylor.
\newblock Modeling stochastic volatility: A review and comparative study.
\newblock {\em Mathematical finance}, 4(2):183--204, 1994.

\bibitem{Breidt1998}
F~Jay Breidt, Nuno Crato, and Pedro De~Lima.
\newblock The detection and estimation of long memory in stochastic volatility.
\newblock {\em Journal of econometrics}, 83(1-2):325--348, 1998.

\bibitem{Singh2016}
Ajay Singh and Dinghai Xu.
\newblock Random matrix application to correlations amongst the volatility of
  assets.
\newblock {\em Quantitative Finance}, 16(1):69--83, 2016.

\bibitem{Laloux1999}
Laurent Laloux, Pierre Cizeau, Jean-Philippe Bouchaud, and Marc Potters.
\newblock Noise dressing of financial correlation matrices.
\newblock {\em Physical review letters}, 83(7):1467, 1999.

\bibitem{Borghesi2007}
Christian Borghesi, Matteo Marsili, and Salvatore Miccich{\`e}.
\newblock Emergence of time-horizon invariant correlation structure in
  financial returns by subtraction of the market mode.
\newblock {\em Physical Review E}, 76(2):026104, 2007.

\bibitem{Musmeci2016}
T.~Di~Matteo N.~Musmeci, T.~Aste.
\newblock Interplay between past market correlation structure changes and
  future volatility outbursts.
\newblock {\em Scientific Reports 6}, 6:36320, 2016.

\bibitem{Zou2005}
Hui Zou and Trevor Hastie.
\newblock Regularization and variable selection via the elastic net.
\newblock {\em Journal of the Royal Statistical Society: Series B (Statistical
  Methodology)}, 67(2):301--320, 2005.

\bibitem{Cont2001}
Rama Cont.
\newblock Empirical properties of asset returns: stylized facts and statistical
  issues.
\newblock {\em Quantitative Finance}, 2001.

\bibitem{Chakraborti2011}
Anirban Chakraborti, Ioane~Muni Toke, Marco Patriarca, and Fr{\'e}d{\'e}ric
  Abergel.
\newblock Econophysics review: Ii. agent-based models.
\newblock {\em Quantitative Finance}, 11(7):1013--1041, 2011.

\bibitem{Mandelbrot1997}
Benoit~B Mandelbrot.
\newblock The variation of certain speculative prices.
\newblock In {\em Fractals and Scaling in Finance}, pages 371--418. Springer,
  1997.

\bibitem{Theil1992}
Henri Theil.
\newblock A rank-invariant method of linear and polynomial regression analysis.
\newblock In {\em Henri Theil’s contributions to economics and econometrics},
  pages 345--381. Springer, 1992.

\bibitem{Micciche2013}
S~Micciche.
\newblock Empirical relationship between stocks cross-correlation and stocks
  volatility clustering.
\newblock {\em Journal of Statistical Mechanics: Theory and Experiment},
  2013(05):P05015, 2013.

\bibitem{Box2015}
George~EP Box, Gwilym~M Jenkins, Gregory~C Reinsel, and Greta~M Ljung.
\newblock {\em Time series analysis: forecasting and control}, page~33.
\newblock John Wiley \& Sons, 2015.

\bibitem{Sachs2012}
Lothar Sachs.
\newblock {\em Applied statistics: a handbook of techniques}.
\newblock Springer Science \& Business Media, 2012.

\bibitem{Musmeci2015b}
Nicolo Musmeci, Tomaso Aste, and Tiziana Di~Matteo.
\newblock Risk diversification: a study of persistence with a filtered
  correlation-network approach.
\newblock {\em Journal of Network Theory in Finance}, 1(1):77--98, 2015.

\bibitem{Tumminello2011}
Michele Tumminello, Salvatore Micciche, Fabrizio Lillo, Jyrki Piilo, and
  Rosario~N Mantegna.
\newblock Statistically validated networks in bipartite complex systems.
\newblock {\em PloS one}, 6(3):e17994, 2011.

\bibitem{Feller2008}
Willliam Feller.
\newblock {\em An introduction to probability theory and its applications},
  volume~2.
\newblock John Wiley \& Sons, 2008.

\bibitem{Jolliffe1982}
Ian~T Jolliffe.
\newblock A note on the use of principal components in regression.
\newblock {\em Applied Statistics}, pages 300--303, 1982.

\bibitem{Preacher2013}
Kristopher~J Preacher, Guangjian Zhang, Cheongtag Kim, and Gerhard Mels.
\newblock Choosing the optimal number of factors in exploratory factor
  analysis: A model selection perspective.
\newblock {\em Multivariate Behavioral Research}, 48(1):28--56, 2013.

\bibitem{Child2006}
Dennis Child.
\newblock {\em The essentials of factor analysis}.
\newblock A\&C Black, 2006.

\bibitem{Lockhart2014}
Richard Lockhart, Jonathan Taylor, Ryan~J Tibshirani, and Robert Tibshirani.
\newblock A significance test for the lasso.
\newblock {\em Annals of statistics}, 42(2):413, 2014.

\end{thebibliography}
\end{document}